\documentclass[prb , superscriptaddress, showpacs, floatfix, nofootinbib, svgnames, eqsecnum, eprint]{revtex4-2}

\usepackage{graphicx}
\usepackage {physics}
\usepackage{tikz}
\usetikzlibrary{shapes.geometric} 
\usepackage{footnote}
\usepackage{hyperref}
\usepackage[normalem]{ulem}

\hypersetup{
    colorlinks = true,
    linkcolor = Red,     
    urlcolor = Green ,
    citecolor = Red ,
    }
    
\usepackage{bm}
\usepackage[caption = false]{subfig}

\tikzstyle{startstop} = [rectangle, rounded corners, 
minimum width=1.5cm, 
minimum height=0.7cm,
text centered, 
draw=Green, 
thick]

\tikzstyle{process} = [rectangle, 
minimum width=1cm, 
minimum height=0.7cm, 
text centered, 
draw=Green, 
thick]

\tikzstyle{decision} = [diamond, 
minimum width=1cm, 
minimum height=0.5cm, 
text centered, 
draw=Green,
thick]

\tikzstyle{arrow} = [thick,->,>=stealth, draw = Green]

\begin{document}

\title{Strong Local Bosonic Fluctuation: The Key to Understanding Strongly Correlated Metals }

\author {S.R.Hassan}
\email {shassan@imsc.res.in}
\affiliation {Institute of Mathematical Sciences, CIT Campus, Tharamani, Chennai 600113, India}
\affiliation {Homi Bhabha National Institute, Training School Complex, Anushakti Nagar, Mumbai
400085, India}

\author{Gopal Prakash}
\email {gopalp@imsc.res.in}
\affiliation {Institute of Mathematical Sciences, CIT Campus, Tharamani, Chennai 600113, India}
\affiliation {Homi Bhabha National Institute, Training School Complex, Anushakti Nagar, Mumbai
400085, India}

\author{N.S.Vidhyadhiraja}
\email {raja@jncasr.ac.in}
\affiliation{JNCASR, Jakkur P.O., Bangalore, India}

\author{T.V.Ramakrishnan}
\email {tvrama2002@yahoo.co.in}
\affiliation {Department of Physics, Indian Institute of Science, Bangalore, India}
\affiliation{JNCASR, Jakkur P.O., Bangalore, India}

\begin{abstract} 

In this paper, we present a theoretical framework for understanding the Extremely Correlated Fermi Liquid (ECFL) phenomenon within the 
$U=\infty$ Hubbard model. Our approach involves deriving equations of motion for the single-particle Green's function 
$G$ and its associated self-energy $\Sigma$, which involves the product of 
the bosonic correlation function comprising both density ($D_N$) and spin ($D_S$) correlations with 
$G$. By solving these equations self-consistently, we explore the behavior of $G$, $D_N$, and 
$D_S$ as functions of frequency, temperature, and hole concentration. Our results reveal distinct coherent and 
incoherent Fermi liquid regimes characterized by the presence or absence of quasiparticle excitations. Additionally, we analyze the intrinsic dc resistivity 
$\rho(T)$, observing a crossover  from $T^2$ to linear behavior with increasing temperature. Our findings delineate Fermi liquid, quantum incoherent, and `classical' regimes in strongly correlated systems, emphasizing the importance of quantum diffusive local charge and spin fluctuations.
\end{abstract}

\maketitle


\section{Introduction}
\label {section 1}

Strongly correlated electronic systems are a fertile ground for phenomena that are unusual, unexpected, and often poorly understood, such as high-temperature superconductivity, metallic states without quasiparticles, and resistivity that varies linearly with temperature over a wide range. Traditional theories, e.g. those based on well defined quasiparticles with the concept of adiabatic continuity between the Drude or free electron gas models and interacting many-electron systems, have been successful in many contexts but fall short in providing qualitative explanations for these phenomena.

The Hubbard model, introduced by Hubbard (\cite{Hubbard1}) and reviewed in a 2022 analysis by Arovas et al. (\cite{Arovas}), adeptly captures the nuances of local electronic interactions. In a simplified scenario where each lattice site in a homogeneous system hosts one orbital, the model is defined by an energy $\epsilon$ at each site $i$, the intersite hopping amplitude $t_{ij}$, and the local correlation energy $U$. The latter represents the additional energy cost for accommodating two electrons with opposite spins at the same site, and is the critical parameter for electron interaction effects. For small and intermediate values of $(U/t)$, where $t$ is the nearest neighbour hopping amplitude, there exists a continuum of (U/t) values 
with the solvable Drude limit of the free electron gas at $(U/t)=0$, allowing for well-defined quasiparticles. However, as $U$ increases, the system crosses over through a Mott metal-insulator phase change for commensurate electron densities
, a phenomenon not accounted for in the quasiparticle framework, heralding the onset of a strongly correlated regime that has been the focus of intensive research for over fifty years. This discussion centers on the paradigmatic limit of strong correlation, specifically $U/t=\infty$, known as the extremely correlated Fermi liquid (ECFL) phase as proposed by Shastry, to derive results that resonate broadly with the characteristics of strongly correlated systems. At both $U=0$ and $U=\infty$ limits (representing the free gas and ECFL, respectively), the carrier density, indicated here by the hole density $\delta$, is 
the sole material parameter. For large but finite $(U/t)$, the small parameter relative to this limit becomes $(t/U)$, leading to the emergence of symmetry-broken states such as antiferromagnetism and superconductivity.

In the case of cuprates, for instance, both electronic structure calculations and experimental findings suggest $U \approx 4eV$ and $t \approx 0.4eV$ (\cite{Sheshadri}), rendering $(U/t) \approx 10 \gg 1$ and thereby positioning the $U/t \rightarrow \infty$ limit as a natural analytical starting point. The literature on the strong correlation problem is extensive; however, a few directions stand out. For infinitely strong correlation, states with local double occupation are excluded via a site-local Gutzwiller projection operator, resulting in `projected' noncanonical fermions. Among the numerous studies on this approach, notable are a 2009 review by Gebhard and Gutzwiller (\cite{Gebhard}) and a 2007 review on a strong correlation (RVB) theory of superconductivity employing the Gutzwiller projection extensively (\cite{Muthukumar}). Another strategy employs a faithful representation of local states via Hubbard $X$ operators (\cite{Hubbard2}), a method extensively elucidated by Ovchinnikov and Val'kov (\cite{Ovchinnikov}). This representation, where Fermi-like and Bose-like $X$ fields (lattice fields) deviate from canonical Fermi or Bose fields, has been widely adopted. Within this framework, Shastry and colleagues have developed a substantial body of work using the Schwinger source method, notably in the $U=\infty$ limit. This well-known approach is described in the condensed matter context, for example, by Baym and Kadanoff (\cite{Baym}), and by Tremblay (\cite{Tremblaybook}), with the foundational paper by Shastry serving as a reference (\cite{Shastry1}), alongside an annotated list (\cite{Shastry2}) and a recent publication (\cite{Shastry3}). Furthermore, many auxiliary field theories have been developed to describe the correlated Fermi system with finite $U$ using canonical fermionic and bosonic fields subject to local constraints. These theories imply that the involved basic quantum fields are not canonical Fermi or Bose fields, but have additional local constraints which are generally applied globally (\cite{Wolfle}). The exploration of metals with $(U/t) >> 1$, namely the strongly correlated Hubbard model metal, spans more than half a century.

Against this backdrop, we develop a simple, approximate, self-consistent theory for the $(U/t) =\infty$ system or the ECFL, employing the equation of motion approach for $G$, the single-particle Green's function. The Dysonian self-energy $\Sigma$ incorporates local charge and spin fluctuation correlators ($D_{N}$ and $D_{S}$ respectively). The equation of motion for these correlators leads to an equation involving the current-current correlator, which in turn involves only a product of two $G$'s for large $d$. By numerically solving the resulting coupled equations, we identify and self-consistently determine $G$ and thereby other physical quantities. We uncover two novel generic features of strongly correlated matter: a smooth crossover from a low-temperature coherent Fermi liquid with well-defined quasiparticles to an incoherent high-temperature Fermi liquid lacking quasiparticles, marked by intrinsic linear resistivity at `high' temperatures.

This paper is organized as follows: Section \ref{section 2} discusses the developed approximate theory. Subsequent sections detail the results from the self-consistent solution of the coupled equations for $G$ and $D$, focusing on the same-site bosonic correlation functions (charge, spin, and current) (Section \ref{section3}), $\Sigma$, and dc resistivity (Section \ref{section4}). The concluding Section (Section \ref{section 5} outlines some limitations and future research directions. Appendices A to D provide in-depth information on several results referenced in the text, including a description of the $X$ operators (Appendix \ref{Appendix A}), the charge and spin correlation function and their equations of motion (Appendix \ref{Appendix B}), the current-current correlation for charge and spin (Appendix \ref{Appendix C}), and an analytical demonstration of coherent Fermi liquid-like behavior at low temperatures (Appendix \ref{AppendixD}).

Our approach to the $(U/t)=\infty$ limit, within a simplified materials-oriented model, identifies the universal origins of two prominent characteristics of strongly correlated systems: the evolution from coherent to incoherent Fermi liquid states with temperature, and linear temperature-dependent resistivity. These phenomena stem from diffusive fluctuations of local electron number (charge) and spin, with the inevitable coupling to constituent Fermi-like excitations shaping electron dynamics. At high temperatures, this interaction manifests as thermal classical electrical noise (white, Johnson-Nyquist noise) at each lattice site, exhibiting a universal amplitude proportional to $T$. This perspective diverges from models that seek to derive observed physical properties (e.g., the strange metal behavior) from a theory of canonical fermions coupled to quantum critical (canonical) Bose fields which avoids the assumption of such fields or their quantum criticality; this typically confines the analysis to a specific point in parameter space.

 
\section{Theory:}
\label{section 2}

We introduce an approximate, self-consistent framework for analyzing the electronic characteristics of metals under the condition of maximal correlation strength. Drawing inspiration from the work of Plakida and associates \cite{Plakida1}, our approach incorporates several novel elements: the use of a high-dimensional ($d$) model, an innovative technique for the calculation of local charge and spin correlation functions, a self-consistent approach to solving relevant connected equations, and the application of the extreme correlation scenario where $(U/t) = \infty$.

Our procedure starts 
with the calculation of the retarded Green's function (GF) for the fermionic $X$-operator, achieved by formulating its equation of motion through differentiation with respect to time variables $t$ and $t'$. This formulation yields an expression that includes a thermal average of four $X$ fields, consisting of both Bose-like and Fermi-like pairs. As $d$ becomes large, the expression simplifies, retaining only same-site Bose-like and Fermi-like fields, interconnected through electron hopping. Approximating in a manner similar to the non-crossing approximation (NCA), we reduce this average to products of local bosonic ($D_N$ and $D_S$) and fermionic correlation functions, subsequently linked to the Dysonian self-energy $\Sigma$ of the initial GF. We then derive equations of motion for the bosonic electron number and spin correlation functions, connecting their time derivatives to current-current correlation functions, which act as a sort of `self-energy'. In the limit of large $d$, this `self-energy' is depicted as the product of two fermionic GFs, culminating in a set of self-consistently solved coupled equations for $G$ and $D$. This methodology facilitates the direct determination of electronic properties such as charge and spin correlation functions, alongside dc and ac electrical resistivities, via $G$ in the high-dimensional limit.

The Hubbard Hamiltonian for the $U = \infty$ scenario, adopting the precise $X$ operator representation for physical quantities as detailed in Appendix \ref{Appendix A}, (assuming energy levels $\epsilon_{1}^\sigma$ for a single particle state with spin $\sigma$ and $\epsilon_{0}$ for the zero particle state are equal (zero), with the system's chemical potential denoted as $\mu$) is: 

\begin{equation}
H = - \mu \sum_{i,\sigma} X_{i}^{\sigma \sigma} + \sum_{ij} t_{ij} X_{i}^{\sigma 0} X_{j}^{0 \sigma},
\end{equation}

where $X_{i}^{\sigma \sigma}$ symbolizes the number operator, $t_{ij}$ indicates the electron hopping matrix element between sites $i$ and $j$, and $X_{i}^{\sigma 0}$ serves as a creation operator, introducing an electron with spin $\sigma$ at site $i$ initially free of electrons.
The focus is on the double-time, retarded Green's function for the fermionic $X$-operator, defined as
\begin{equation}
    \begin{aligned}
    G_{ij}^{R \sigma\sigma^{\prime}}(t,t^{\prime}) & = -i\theta(t-t^\prime) \expval{ \comm {X_{i}^{0\sigma}(t)} {X_{j}^{\sigma^{\prime}0}(t^{\prime})}_{+}} \equiv \expval{\expval{X_{i}^{0\sigma}(t) \big | {X_{j}^{\sigma^{\prime}0}(t^{\prime})}}}
    \end{aligned}
    \label{eqn2T}
\end{equation}
where $\comm{~}{~}_{+}$ denotes the anticommutator and $\langle...\rangle$ represents the expectation value in the grand canonical ensemble. The second notation on the right-hand side offers a more concise depiction of the first term.

To derive $G_{ij}^{R \sigma\sigma^{\prime}}(t,t^{\prime})$, we utilize the equation of motion method.\footnote{Henceforth, we focus on the spin diagonal case $\sigma = \sigma^{\prime}$, as only this configuration is non-zero for paramagnetic spin isotropic systems with spin quantization along the $z$ axis; additionally, $G$ is independent of $\sigma$.} The equation of motion for the Green's function defined in equation (\ref{eqn2T}) is
\begin{align}
   i\partial_{t}G_{ij}^{R\sigma\sigma}(t,t^{\prime})&=  Q\delta_{ij}\delta(t-t^{\prime})-\mu G_{ij}^{R\sigma\sigma}(t,t^{\prime})
+Q\sum_{l}t_{il}G_{lj}^{R\sigma\sigma}(t,t^{\prime})
   +{\mathcal{G}}_{ij}^{R\sigma \sigma}(t,t^{\prime})\mbox{,}
   \label{eqn3T}
\end{align}
where $Q= \ev{B_{i}^{\sigma\sigma}} = \overline{B_{i}^{\sigma\sigma}}$ and ${\mathcal G}_{ij}^{R\sigma\sigma}(t,t^{\prime})$ is a higher-order Green's function (originating from fluctuations in the Bosonic operator $B_{i}^{\sigma \sigma^{\prime}}(t)$) and is detailed as:
\begin{align}
    {\mathcal G}_{ij}^{R\sigma\sigma}(t,t^{\prime}) & = \sum_{l\sigma^{\prime \prime}}t_{il} 
    \expval{\expval{\delta  B_{i}^{\sigma \sigma^{\prime \prime}}(t)X_{l}^{0\sigma^{\prime \prime}}(t) \big| X_{j}^{\sigma 0}(t^{\prime})}} \mbox{.}
    \label{eqn4T}\\
{\rm where}\;\;
\delta{B}_{i}^{\sigma\sigma^\prime}(t) &= {B}_{i}^{\sigma\sigma^\prime}(t)- \ev{{B}_{i}^{\sigma\sigma^\prime}(t)}\,.
    \label{eqn5T}
\end{align}
As elucidated in Appendix \ref{Appendix A},
$B_{i}^{\sigma \sigma^{\prime}} = \delta_{\sigma\sigma^{\prime}} X_{i}^{00} + X_{i}^{\sigma^{\prime}\sigma}$. This represents a local charge operator for $\sigma = \sigma^\prime$ and a local spin flip for $\sigma \neq \sigma^\prime$.

The equation of motion, Eq.(\ref{eqn3T}), when expressed in frequency space (under equilibrium conditions), transforms to:
\begin{align}
G^{R}_{ij}(\omega)=G_{ij}^{R,MF}(\omega)+\sum_{l}G_{il}^{R,MF}(\omega)\frac{1}{Q}{\mathcal G}^{R}_{lj}(\omega)
\label{eqn6T}
\end{align}
where the mean-field Green's function $G_{ij}^{R,MF}$ is outlined as:
\begin{equation}
G_{ij}^{R,MF} =\sum e^{ik\cdot(R_i-R_j)}G_{k}^{R,MF}(\omega)
{\rm \quad and \quad}
G_{k}^{R,MF}(\omega)=
\frac{Q}{(\omega+\mu - Q\epsilon_{k}+i0^{+})},
\label{eqn7T}
\end{equation}
Here, $\epsilon(\vec k) = -2t(\cos{k_x}+\cos{k_y})$ exemplifies a two-dimensional scenario, for example, in the case of a square planar lattice.

The fluctuation component ${\mathcal G}_{ij}^{R\sigma\sigma}(\omega)$, obtained through the Fourier transform (FT) of ${\mathcal G}_{ij}^{R\sigma\sigma}(t-t')$ as defined in equation (\ref{eqn4T}), is determined via its equation of motion concerning $t^\prime$. In frequency space, the resultant expression is:

\begin{align}
    (\omega +\mu) {\mathcal{G}}_{ij}^{R \sigma \sigma} & =  \sum_{l \sigma''} t_{il}\expval{\left[\delta B_i^{\sigma\sigma''}X_{l}^{0\sigma''}, X_{j}^{\sigma 0}\right]_+} + Q \sum_{l} t_{il} {\mathcal{G}}_{lj}^{R \sigma \sigma}(\omega) \nonumber \\ 
    + & \qty ( \sum_{l l^\prime \sigma^{\prime \prime}\sigma^{\prime \prime \prime}}t_{l^{\prime}j}t_{il} \expval{\expval{\delta{B}_{i}^{\sigma \sigma^{\prime \prime}}(t) X_{l}^{0\sigma^{\prime \prime}}(t) \big| {{\delta B}_{j}}^{\sigma \sigma^{\prime \prime \prime}}(t^{\prime})X_{l^{\prime}}^{\sigma^{\prime \prime \prime} 0}(t^{\prime})}}_{\omega} )_{FT}
    \label{eqn8T}
\end{align}
The  first term on the right in the above equation vanishes by design, as described in the projection operator formalism of Plakida (\cite{Plakida1}). Explicitly, one has

\begin{equation}
\sum_{l\sigma''} t_{il} \ev{{\comm{\delta B_{i}^{\sigma\sigma''}(t)X_{l}^{0\sigma''}(t)}{X_{j}^{\sigma0}(t)}}_{+}}=0 .
    \label{eqn9T}
\end{equation}
This condition can be thought of in the Mori- Zwanzig memory function language (see e.g. ref.(\cite{Mori}) and the book by Forster (\cite{Forster}) as related to the `noise' implied in their Liouvillean operator projection scheme. It turns out in the present case that $\delta{B}_{i}^{\sigma\sigma^\prime}(t)$ defined this way is the same as in Eq.(\ref{eqn5T}).

After applying a spatial Fourier transform to Eq. (\ref{eqn8T}) we obtain:

\begin{equation}
    {\mathcal G}^{R}_{k}(\omega) = T_{k}(\omega) \frac{1}{Q}G^{R,MF}_{k}(\omega)
    \label{eqn10T}\,.
\end{equation}

In this equation, the superscripts $\sigma\sigma$ are omitted, and $T$ is identified as a scattering matrix defined by:
\begin{align}
    T_{k}^{\sigma\sigma}(\omega) =  \qty ( \sum_{l l^\prime \sigma^{\prime \prime}\sigma^{\prime \prime \prime}}t_{l^{\prime}j}t_{il} \expval{\expval{\delta{B}_{i}^{\sigma \sigma^{\prime \prime}}(t) X_{l}^{0\sigma^{\prime \prime}}(t) \big| {{\delta B}_{j}}^{\sigma \sigma^{\prime \prime \prime}}(t^{\prime})X_{l^{\prime}}^{\sigma^{\prime \prime \prime} 0}(t^{\prime})}}_{\omega} )_{FT}.
    \label{eqn11T}
\end{align}
Implementing equation (\ref{eqn10T}) in equation (\ref{eqn6T}) results in an equation for $G$ as:

\begin{align}
     G = G^{MF} + G^{MF}{\tilde T}G^{MF}\hspace{0.1cm}\mbox{, with the scattering matrix}\hspace{0.2cm} {\tilde T} = \frac{T}{Q^2}\,.
\label{eqn12T}
\end{align}
This equation, expressed in Dyson's form, becomes:
\begin{equation}
G=G^{MF}+G^{MF}\Sigma G,
\label{eqn13T}
\end{equation}

The Dysonian self-energy $\Sigma$ is related to ${\tilde T}$ through:
\begin{align}
{\tilde T} = \Sigma + \Sigma\, G^{MF}\, {\tilde T}\,.
\label{eqn14T}
\end{align}
The self-energy $\Sigma$ adopts the same averaged form as $\tilde{T}$. We propose:
\begin{equation}
\Sigma_{ij}^{R,\sigma\sigma}(\omega) =  \frac{1}{Q^2} \sum_{l l^\prime \sigma^{\prime \prime}\sigma^{\prime \prime \prime}}t_{l^{\prime}j}t_{il} {\expval{\expval{ B_{i}^{\sigma \sigma^{\prime \prime}}(t)X_{l}^{0\sigma^{\prime \prime}}(t) \big| {{B_{j}}^{\dagger}}^{\sigma \sigma^{\prime \prime \prime}}(t^{\prime})X_{l^{\prime}}^{\sigma^{\prime \prime \prime} 0}(t^{\prime})} }}_{\omega}\,.
\label{eqn15T}
\end{equation}

indicating that $\Sigma({B})$ corresponds to $\tilde T(\delta B)$, particularly at high frequencies. The self energy term above, under a Bose Fermi or DG decoupling is the same as the self-consistent Born approximation (SCBA) discussed by Plakida (see e.g. ref.\cite{Plakida1}, \cite{Plakida2}) , or the non-crossing approximation (NCA). This form closely resembles Hubbard's description \cite{Hubbard3} of the leading `scattering correction' term or self-energy, extending beyond the mean field term presented in Eq.(\ref{eqn7T}).

In the large $d$ approximation, the self-energy becomes site-local ($i=j$), predominantly influenced by the nearest neighbour $l=l'$ to site $i$. With the only non-zero hopping term being $t_{il}= t_{}/\sqrt{d}$ (where $t_{} =1$), the summation over $l$ introduces a factor of $d$. To calculate the self-energy, a decoupling approximation is applied as follows:

\begin{align}
    \Sigma_{ii}^{R,\sigma\sigma}(\omega) = \frac{1}{Q^{2}} \Bigg[ -i \theta(t-t') \Bigg( \sum_{\sigma''} \ev{B_{i}^{\sigma \sigma''}(t) {{B_{i}}^{\dagger}}^{\sigma \sigma''}(t')} \ev{X_{i}^{0 \sigma''}(t) X_{i}^{\sigma'' 0}(t') } \nonumber
    \\+ \ev{{{B_{i}}^{\dagger}}^{\sigma \sigma''}(t') B_{i}^{\sigma \sigma''}(t)} \ev{ X_{i}^{\sigma'' 0}(t') X_{i}^{0 \sigma''}(t)} \Bigg)\Bigg]_{\omega}  
    \label{eqn16T} 
\end{align}
The approximation error in this decoupling is also of relative order $(1/d)$.

Utilizing the spectral representation for the correlation functions specified above and connecting them to the spectral representation of the retarded Green's functions, the local self-energy $\Sigma^{R}(\omega)$ is formulated as:

\begin{align}
\Sigma^{R}(\omega) & = -\frac{1}{Q^2}\int_{-\infty}^{\infty} \dd \epsilon_{1}  \dd \epsilon_{2} ~ \rho_{\scriptscriptstyle G}(\epsilon_{1}) \rho_{\scriptscriptstyle D}(\epsilon_{2}) ~ \qty ({\frac{\tanh{(\frac{\beta \epsilon_{1}}{2})}+\coth{(\frac{\beta \epsilon_{2}}{2})}}{\omega^{+}-\epsilon_1-\epsilon_2}}) \label{eqn17T}
\\ {\rm where,}\; \rho_{\scriptscriptstyle G}(\epsilon_1)& =-\frac{1}{\pi}{\Im G^{R}}(\epsilon_1),\;\;\rho_{\scriptscriptstyle D}(\epsilon_2)= - \frac{1}{\pi} \Im D^{R}(\epsilon_2)\,\label{eqn18T} \\
{\rm and,} \;   D^{R} (t,t^{\prime})& =\sum_{\sigma''} \expval{\expval{B^{\sigma \sigma''}(t)| {B^{\dagger}}^{\sigma \sigma''}(t^{\prime})}}= -i \theta(t-t') \sum_{\sigma''}\expval{{\comm{B^{\sigma \sigma''}(t)}{{B^{\dagger}}^{\sigma \sigma''}(t^{\prime})}}_{-}} \,. \label{eqn19T}
\end{align}
$D^{R}$, when expressed in terms of the number $N$ and spin $S_{z}$ operators, is outlined as 
(refer to Appendix \ref{Appendix B} for details):
\begin{align}
D^{R}(t,t^{\prime}) = \frac{1}{4}\qty{-i \theta(t-t') \ev{\big[N(t),N(t^\prime)\big]_{-}}} + \frac{3}{4} \qty{-i \theta(t-t') \ev{ \big[S^{+}(t),S^{-}(t^\prime)\big]_{-}}}
\label{eqn20T}.
\end{align}
Given that computing $D^{R}_{N/S}(\omega)$ through an equation of motion approach is impractical due to the commuting nature of operators, resulting in the disappearance of equal time 
inhomogeneous terms \cite{Zubarev} multiplying the delta function $\delta(t-t')$, alternative strategies are necessary.
 
We adopt a specific approach to compute these functions.
The expressions for $D^{R}_N(t-t^\prime)=\langle\langle N(t) | N(t^\prime)\rangle\rangle$ and $D^{R}_{S}(t-t^\prime)=\langle\langle S^{+}(t) | S^{-}(t^\prime)\rangle\rangle$ are defined as:
\begin{align}
D^{R}_{N}(t-t^\prime) & =D^{+}_{N}(t-t^\prime)+D^{-}_{N}(t-t^\prime) \nonumber \\
&= -i\theta(t-t^\prime) \langle N(t)N(t^\prime)\rangle -i\theta(t-t^\prime)\langle N(t^\prime)N(t)\rangle,
\label{eqn23T} \\
{\rm and,}\; D^{R}_{S}(t-t^\prime) & =D^{+}_{S}(t-t^\prime)+D^{-}_{S}(t-t^\prime)\nonumber\\
&= -i\theta(t-t^\prime)\langle S^{+}(t)S^{-}(t^\prime)\rangle -i\theta(t-t^\prime)\langle S^{-}(t^\prime)S^{+}(t)\rangle \,.
\label{eqn24T}
\end{align}

\begin{figure}
\begin{tikzpicture}[node distance=2cm]
\node (start) [startstop] { $\bm{\Sigma^{R} (\omega) = 0}$ } ;
\node (step1) [process, below of = start, yshift = 0.9cm] { $\bm{G^{R}(\omega)}$ } ;
\node (step2) [process, below of = step1, yshift = 0.9cm] {$\bm{\chi^{R}(\omega) \rightarrow\chi^{+}_{N/S}(\omega)\rightarrow D^{+}(\omega)}$ };
\node (step3) [process , below of = step2, yshift = 0.8cm] {$ \bm{D^{+}(\omega), G^{R}(\omega)\rightarrow\Sigma^{R} (\omega)}$};
\node (check) [decision, below of = step3,yshift = 0.3cm] {\textbf{conv?}};
\node (feedback) [process, right of= check, xshift=1cm] {\textbf{updated} $\bm{\Sigma^{R} (\omega)}$};
\node (stop) [startstop, below of = check, yshift = 0.3cm] {\textbf{output, stop}};
\draw [arrow] (start) -- (step1);
\draw [arrow] (step1) -- (step2);
\draw [arrow] (step2) -- (step3);
\draw [arrow] (step3) -- (check);
\draw [arrow] (check) -- node[anchor = east]{yes} (stop);
\draw [arrow] (check) -- node[anchor = south]{no} (feedback);
\draw [arrow] (feedback) |- (step1);
\end{tikzpicture}
    \caption{Self-consistency loop. This is a schematic illustration of the numerical scheme in which the input $G^R{(\omega)}$ and the output $G^R{(\omega)}$ should match for self consistency}
\label {Selfconsistency}
\end{figure}
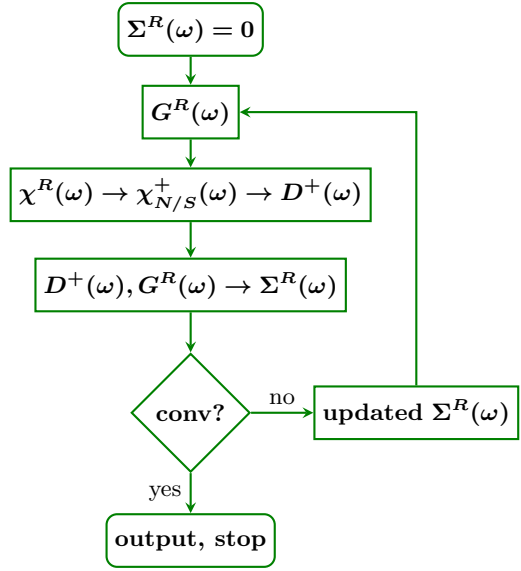

In the context of spectral functions, the formulation is as follows:
\begin{align}
\hspace{-4.8cm} D^{\alpha}_{\gamma}(\omega)=\int_{-\infty}^{\infty} d \omega^{\prime} \frac{\rho^\alpha_{\scriptscriptstyle D_{\scriptscriptstyle \gamma}}(\omega^{\prime})}{\omega-\omega^{\prime}+i0^{+}}
\end{align}
where the indices $\alpha$ can take values of $+$ or $-$, and $\gamma$ represents either $N$ or $S$. The spectral functions, denoted as $\rho^\alpha_{ D_{ \gamma}}=-{\rm Im}(D^\alpha_\gamma)/\pi$, adhere to the following relationships:

\begin{align}
    D^{R}_{\gamma}(\omega)&=\sum_{\alpha}D^{\alpha}_{\gamma}(\omega)\mbox{,} ~ \rho_{\scriptscriptstyle D_{\scriptscriptstyle N/S}}(\omega)=\sum_{\alpha}\rho^{\alpha}_{\scriptscriptstyle D_{\scriptscriptstyle N/S}}(\omega)
\label{eqn27T}\\ 
    \rho_{\scriptscriptstyle D_{\scriptscriptstyle N/S}}(\omega)&=-\rho_{\scriptscriptstyle D_{\scriptscriptstyle N/S}}(-\omega)\mbox{,} ~  \rho^{-}_{\scriptscriptstyle D_{\scriptscriptstyle N/S}}(\omega)=-e^{-\beta \omega}\rho^{+}_{\scriptscriptstyle D_{\scriptscriptstyle N/S}}(\omega)\,, 
\label{eqn28T} 
\end{align}

$\rho^{\alpha}_{\scriptscriptstyle D_{\scriptscriptstyle N}}(\omega)$ satisfies the following sum rule:

\begin{equation}
\int_{-\infty}^{\infty}d\omega\rho^{\alpha}_{\scriptscriptstyle D_{\scriptscriptstyle N}}(\omega)=\alpha n
\label{sumrule}
\end{equation}

Following this, an equation of motion method for $D^{\alpha}_{\gamma}(\omega)$ is established, paralleling the approach above
for $G^{R}(\omega)$ (This is detailed in Appendix \ref{Appendix B}).

This yields the expressions for $D^{\alpha}_{N/S}(\omega)$ as:
\begin{equation}  \left(D^{\alpha}_{\scriptscriptstyle N}(\omega)\right)^{-1}= \alpha\frac{1}{n}\left(\omega-\alpha\frac{\chi^{\alpha}_{\scriptscriptstyle N}(\omega)}{n}\right)\,, \quad
\left(D^{\alpha}_{\scriptscriptstyle S}(\omega)\right)^{-1}= \alpha\frac{2}{n}\left(\omega -\alpha\frac{\chi^{\alpha}_{\scriptscriptstyle S}(\omega)}{\frac{n}{2}}\right)
    \label{eqn30T}
\end{equation}
where $\chi^{\alpha}_{\scriptscriptstyle N/S}(\omega)$ is the Fourier transform of the following:
\begin{align}
    \chi^{+}_{\scriptscriptstyle S}(t)=-i\theta(t)\expval{J_s(t)J_{s}(0)}\mbox{,} \quad
    \chi^{+}_{\scriptscriptstyle N}(t)=-i\theta(t)\expval{J_c(t)J_{c}(0)}
    \\ \chi^{-}_{\scriptscriptstyle S}(t)=-i\theta(t)\expval{J_s(0)J_{s}(t)}\mbox{,} \quad
    \chi^{-}_{\scriptscriptstyle N}(t)=-i\theta(t)\expval{J_c(0)J_{c}(t)}
\label{eqn31T}
\end{align}
Here, $J_s$ and $J_c$ are defined as the spin and charge currents respectively:
\begin{align}
J_{s} = \frac{1}{N}\sum_{k}v_{k}X_{k}^{0 \sigma}X_{k}^{\bar{\sigma}0}, \quad J_{c} = \frac{1}{N}\sum_{k,\sigma}v_{k}X_{k}^{0\sigma}X_{k}^{\sigma0}
\end{align}
where $v_{k} = \partial_{k} \epsilon_{k}$, with $\epsilon_{k}$ being the energy dispersion on the lattice.

The spectral representation for $\chi^{\alpha}_{\gamma}$ is given by:
\begin{align}
    \chi^{\alpha}_{\scriptscriptstyle \gamma}(\omega)&=\int_{-\infty}^{\infty}\dd \omega^{\prime} \frac{\qty{\frac{1+\alpha}{2}+n_B(\omega^\prime)}\rho_{ \scriptscriptstyle \gamma}(\omega^\prime)}{\omega-\omega^\prime+i0^{+}} 
    \label{eqn33T}
     \\ {\rm where,}\;\;\rho_{\scriptscriptstyle \gamma}(\omega^\prime)&=-\frac{1}{\pi}\Im \chi^R_\gamma(\omega^{\prime}) \;\;{\rm and,} \quad n_{B}(\omega^{\prime}) = \frac{1}{e^{\beta \omega^{\prime}} - 1} 
\end{align}
$\chi^{R}(\omega)$ for both spin and charge sectors is derived from the particle-hole bubble diagram, ignoring vertex corrections, as detailed in \cite{DMFTREVIEW}. The relationship $\chi^{R}_{N}(\omega) = 2 \chi^{R}(\omega) $ and $\chi^{R}_{S}(\omega) = \chi^{R}(\omega) $ is outlined in Appendix \ref{Appendix C}, with the current-current correlation in infinite dimensions ($d = \infty$) described as follows:
\begin{align}
\chi^{R}(\omega) = \frac{1}{N} \sum_{k} \int \int d \omega_{1} d \omega_{2} \frac{\rho_{G}(k,\omega_{1}) \rho_{G}(k,\omega_{2}) v_{k}^{2}}{\omega + \omega_{1} - \omega_{2} + i \eta} \left (n_{F}(\omega_{1}) - n_{F}(\omega_{2})\right)
\end{align}
where $\rho_{G}(k,\omega) = -\frac{1}{\pi} \mathrm{Im} G^{R}(k,\omega)$ and $n_{F}(\omega)$ denotes the Fermi function. The calculation of the imaginary part of $\chi^{R}(\omega)$ utilizes the convolution/correlation theorem, with the real part derived from the Kramers-Kronig relation. The adaptation for a Bethe lattice modifies the expression to involve the transport density of states $\Phi(\epsilon)$ \cite{DMFTREVIEW}, leading to a refined calculation of $\mathrm{Im} \chi^{R}(\omega)$ as detailed in the equations.
\begin{align}
    \mathrm{Im} \chi^{R}(\omega) & = - \pi \iint \dd \epsilon ~ \dd \omega_{1} ~\Phi(\epsilon) \rho_{G}(\epsilon,\omega_1) \rho_{G}(\epsilon,\omega + \omega_{1}) \qty {n_{F}(\omega_{1}) - n_{F}(\omega + \omega_{1})}
    \label{curr_corr_exp}
    \\ & \Phi(\epsilon) = \frac{1}{N} \sum_{k} v_{k}^{2} \delta(\epsilon - \epsilon_{k}) = \Phi_{0} (4-\epsilon^{2})^{3/2} \nonumber
\end{align}

The process of determining $\Sigma^{R}, G^{R}, \chi^{R}, D^{R}, D^{\alpha}$, and $\chi^{\alpha}$ involves a self-consistent scheme, which can be summarized as follows:
\begin{enumerate}
    
\item {\bf Initialization}:\\
Begin with an arbitrary selection of $G^R$. Using a specific equation (referred to as Eq.(\ref{curr_corr_exp})), compute $\chi^{R}$ based on the initial $G^R$.
\item {\bf Computation of $\chi^{\alpha}$ and $D^{\alpha}$}: \\
From $\chi^{R}$, calculate $\chi^{\alpha}$ using another equation (Eq.(\ref{eqn33T})). This, in turn, allows for the determination of $D^{\alpha}$ through Eq.(\ref{eqn30T}).
\item {\bf Self-Energy Calculation ($\Sigma^{R}$)}:\\
The self-energy, $\Sigma^{R}$, is calculated using both $D^{R}_{\gamma}$ (from Eq.(\ref{eqn27T})) and the initial or previously computed $G^{R}$ (via Eq.(\ref{eqn17T})).
\item {\bf Update of $G^{R}$}:\\
With the newly computed $\Sigma^{R}$, update the full Green's function, $G^{R}$, using Eq.(\ref{eqn13T}). This updated $G^{R}$ is then used as the starting point for the next iteration of the process.
\end{enumerate}
The cycle repeats, starting from step (1) with the newly obtained $G^{R}$, and continues until $\Sigma^{R}$ converges within a specified tolerance. This iterative procedure, known as the self-consistency loop, ensures that the calculations for $\Sigma^{R}, G^{R}, \chi^{R}, D^{R}, D^{\alpha}$, and $\chi^{\alpha}$ are mutually consistent and converge to a stable solution. The entire self-consistency loop is illustrated in a figure referred to here as Fig.\ref{Selfconsistency}. Throughout this iterative process, adherence to the sum rule, expressed in Eq.(\ref{sumrule}), is maintained.



\section{Local Charge, Spin and Current Correlation Functions:}
\label{section3}
\begin{figure}
\centering
\subfloat[]{\includegraphics[width=0.4\textwidth]{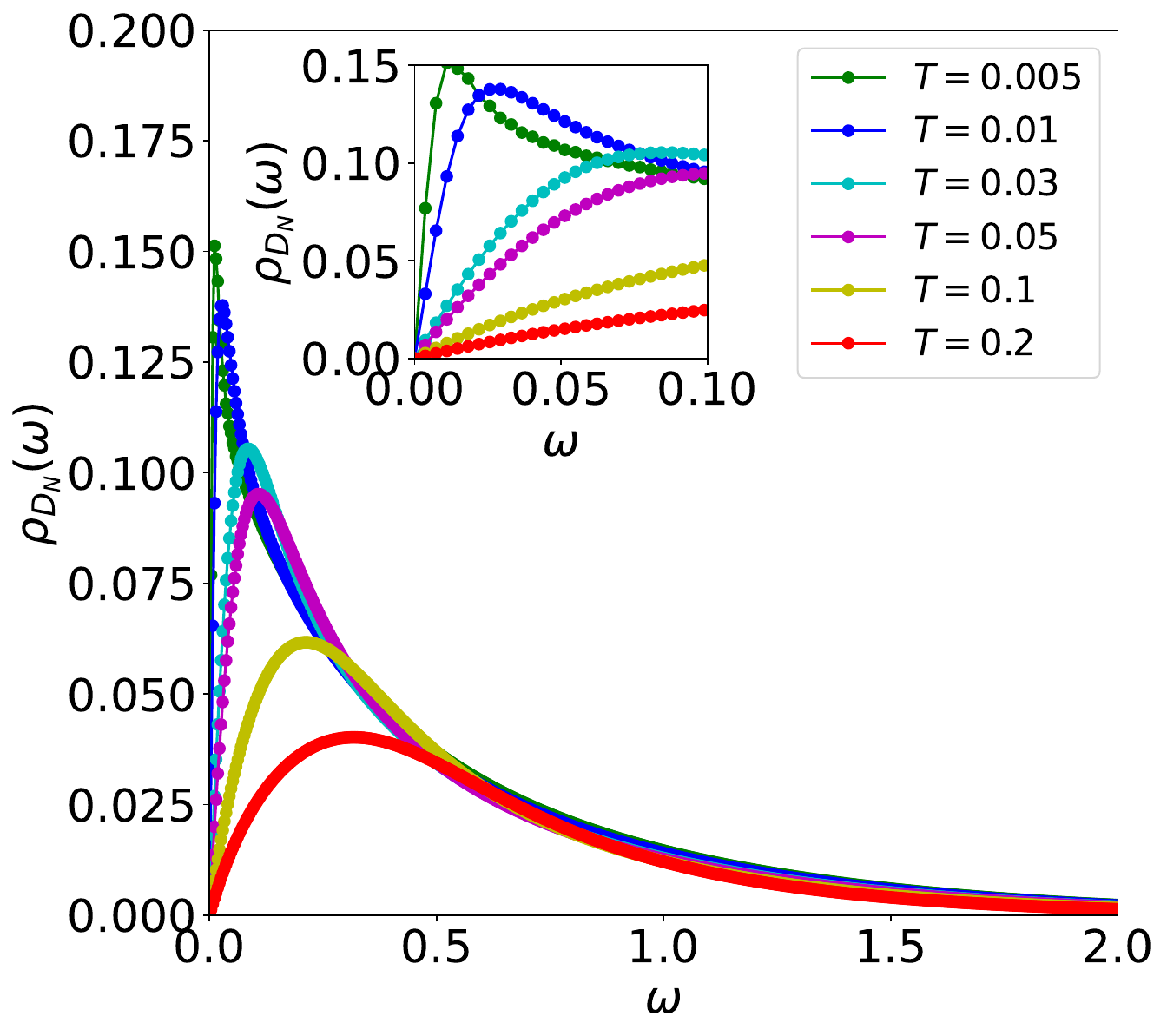}}
\quad
\subfloat[]{\includegraphics[width=0.36\textwidth]{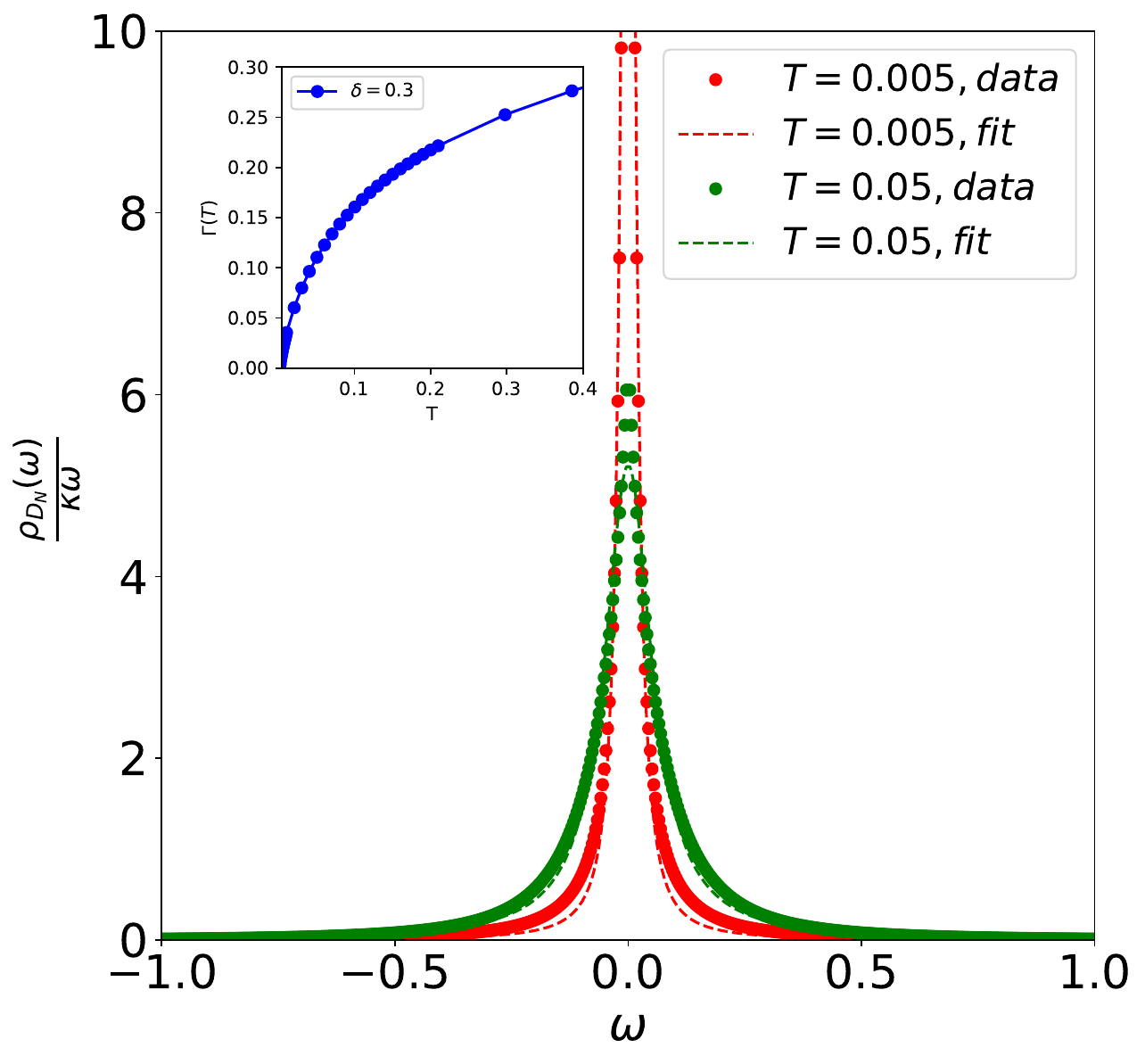}}
\caption{(a) The evolution of $\rho_{D_{N}}(\omega)$ for positive frequency as a function of temperature for doping $\delta=0.3$  
(b) The normalized charge spectral function,$\frac{\rho_{D_N}(\omega)}{\omega\kappa}$, at a fixed temperatures and its fit with Lorentzian; the inset shows the Lorentzian width, $\Gamma(T)$ vs T at $\delta=0.3$}
\label{charge_correlation}
\end{figure}

In this Section, we discuss the bosonic correlation functions mentioned above; these determine the electron dynamics of the infinitely strongly correlated metal, and as discussed in Section \ref{section 2}, they are all related to each other. In the large $d$ limit, charge and spin correlation functions are identical to within numerical factors having to do with the spin $(1/2)$ of the electron. We therefore discuss here only the charge correlation function. We also discuss here the charge current current correlation function which is the `self energy' of the charge correlation function, as seen from Equation ($\ref{eqn30T})$. The real frequency spectral density $\rho_{D_{N}}(\omega)$ of the charge correlation function is shown in Fig $\ref{charge_correlation}$(a) as a function of positive frequency $\omega$ for different temperatures $T$ and at doping $\delta = 0.3$. This has the general properties of being real, positive for positive $\omega$ and antisymmetric with respect to its sign change.

The local charge fluctuation is a massless damped excitation, as is evident from the general shape of $\rho_{D_{N}}(\omega)$ (there is no sharp peak corresponding to a mass term or a restoring force ; on the other hand, its spectral density  has a smooth structure with a generally broad asymmetric peak and a long tail). The charge at each site diffuses quantum mechanically in a time and temperature dependent manner; there is no net restoring force. Since in the electron phonon system, the best known model of a bosonic system coupled to electrons, the quantum scale is set by the Debye frequency  $\omega_{D}$ determined by the nonzero restoring force it is likely that there is no such scale here, and that the occurrence of a small Fermi liquid like regime (section \ref{section4}) is due to a different reason. 

The very existence of such a distinct bosonic fluctuation coupled to electron dynamics is a strong correlation effect, since it is defined in relation to projected fermion or $X$ operator degrees of freedom whose specific properties are determined by strong correlation. Roughly, the diffusion spectrum Fig $\ref{charge_correlation}$(b) consists of a rising part at low frequencies, a peak, and a long tail. The low frequency rise is less steep as temperature increases, as is the fall. Overall, the spectrum can be fitted roughly by a Lorentzian like form going as $\Gamma/[\omega^{2}+ \Gamma^{2} ]$ with $\Gamma$ being the damping constant, as shown in Fig $\ref{charge_correlation}$(b). The actual spectrum decays more rapidly at higher frequencies than this form so that its normalized area is unity, and its first moment is finite. The simplified form is useful since it focuses attention on the quantity $\Gamma$ (see inset of Fig. $\ref{charge_correlation}$(a)) which is the damping rate of fluctuations. It is small at low temperatures, being roughly proportional to $T$ but larger than it and one has well-defined quasiparticles (a Fermi liquid). We discuss later below the implied quantum and classical regimes in local density fluctuations.

\begin{figure}
\centering
\subfloat[]{\includegraphics[width=0.6\textwidth]{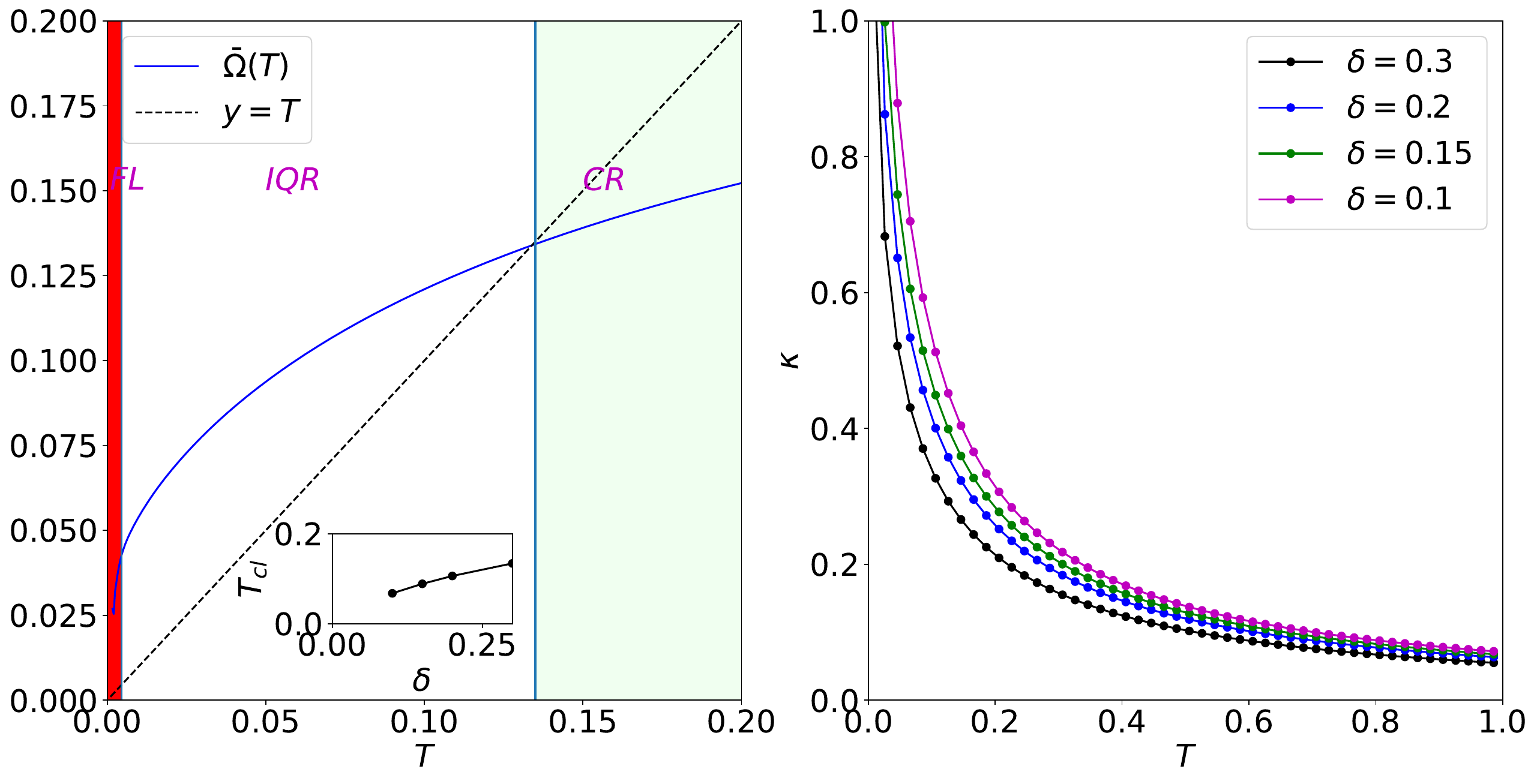}}
\quad
\caption{(a) The average frequency, denoted as $\bar{\Omega(T)}$, varies with temperature for a doping level of $\delta=0.3$. In the temperature range $0<T<0.005$, represented by a shallow red area, the system exhibits Fermi liquid (FL) behavior. For temperatures in the range of $0.005<T<T_{cl}(=0.135)$, the system is within the incoherent Quantum region, crossing over to the classical regime (CR) when $T>T_{cl}$. An inset illustrates the relationship between $T_{cl}$ and doping $\delta$, with a dashed line indicating $Y=T$. (b) shows the charge compressibility, $\kappa$, as a function of temperature $T$ for different doping levels.}

\label{charge_phase_diagram}
\end{figure}
We now infer two consequences of the actual $\rho_{D_{N}}(\omega)$ shown in Fig $\ref{charge_correlation}$(a), one from the low frequency or quantum end, and another from using its overall spread or first moment which weights strongly the higher frequency or classical regime. At low frequencies $\rho_{D_{N}}(\omega)$ (ideally above $\omega = 0$, but in reality above a low nonzero value $\omega_{l} = 0.002t \simeq 8K$ for the large $t = 0.4eV$ ) is seen to be linear in $\omega$ (it is a smooth function and is antisymmetric in $\omega$ so that the leading term near $\omega =0$ has to be linear). As observed from the inset in Fig $\ref{charge_correlation}$(a) for extremely low temperatures, the slope of spectral density ($A(T)$)(not shown in the figure) very close to $\omega = 0$ is almost $T$ independent and this gives rise to canonical Fermi Liquid form of $\Im \Sigma(\omega,T)$(we show that this is true analytically in Appendix $\ref{AppendixD}$).

The typical energy scale of $\rho_{D_{N}}(\omega)$ is the average energy of the local density  fluctuations or the first moment
 \begin{equation}
 \overline{\Omega(T)} = \int_{\omega_{l}}^{\omega_{u}} \omega \qty(\frac{\rho_{D_{N}}(\omega)}{\omega\kappa})\, d\omega \hspace{0.05cm} \mbox{,  where}\hspace{0.15cm} \kappa = \int_{-\infty}^{\infty}\qty(\frac{\rho_{D_{N}}(\omega)}{\omega})d\omega \hspace{0.05cm} \mbox .
 \end{equation}

In the above equation, $\kappa$ is the thermodynamic compressibility, and the upper frequency limit $\omega_{u}$ of the integral is very large but finite; we use $\omega_{u}= 30t$.

We show $\overline{\Omega(T)}$ as a function of $T$ in Fig $\ref{charge_phase_diagram}$(a). It is small at low temperatures, roughly proportional to $T$ but larger than it and one has well-defined quasiparticles (a Fermi liquid). As temperature increases, it increases sublinearly with $T$ and essentially flattens out at high temperatures. The case when $\overline{\Omega(T)}$ is lower than the temperature $T$ defines the classical limit for fluctuations. We observe that for $T > 0.13$ for doping $\delta = 0.3$, we enter into the classical regime. Below $T < 0.13$ we are in a quantum regime and we have a coherent Fermi Liquid phase at extremely low temperatures followed by a linear in $T$ resistivity regime (which also lies in the quantum regime) which is denoted by the Incoherent Quantum Regime (IQR).

The quantities $\Gamma(T)$, $A(T)$ (not shown in the figure) and $\overline{\Omega(T)}$ defined above, are different calculated characteristics of the local charge correlation function describing its diffusion and `stay at home' probability in frequency space. In strongly correlated lattice systems, at `high' temperatures, quantum mechanical intersite hopping $t_{ij}$ can be neglected, and the system is a statistical superposition of energetically degenerate states with one or no charge at a lattice site. This regime is accessed by experiments on thermopower of strongly correlated metals (see e.g. \cite{Ong}) and references in the paper ref. \cite{Subroto}). The thermopower, which measures the entropy of charge carriers, is seen to saturate at values consistent with a Heikes-like estimate of the entropy of this classical metal; experimentally, the `classical' regime begins at surprisingly low temperatures. 

The frequency and momentum dependence of charge fluctuations has been recently explored experimentally using momentum-resolved EELS (see e.g. \cite{Abbamonte}) and RIXS (see e.g. \cite{Arpaia}). They do find essentially nondispersive density fluctuations; namely, they are spatially local, as obtained here. In subsequent work, we will present detailed predictions of this spectrum in our theory.

\begin{figure}
\centering
\includegraphics[scale=0.5]{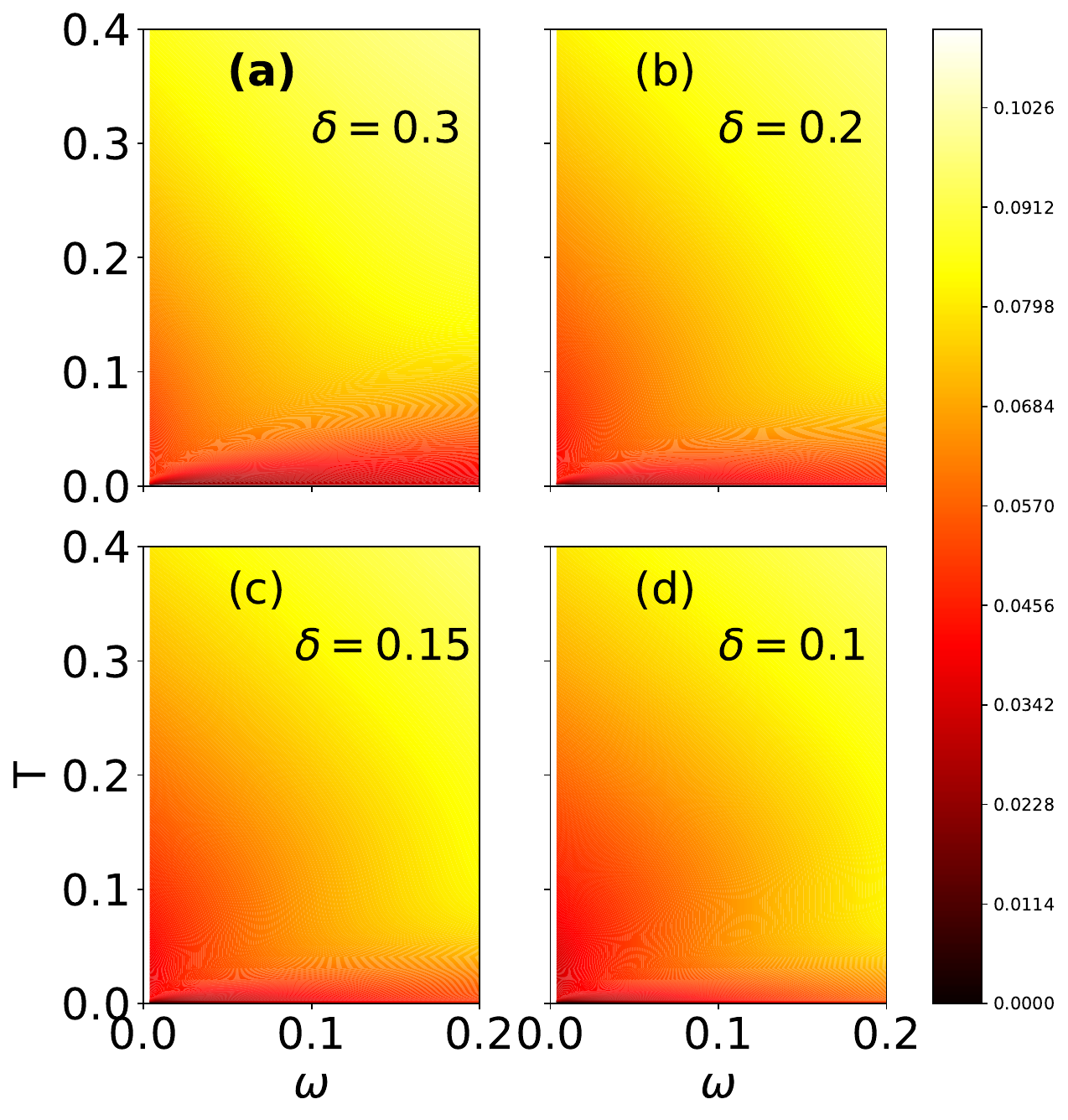}
\caption{Contour plots of imaginary part of current-current correlation function   $\Im \chi^{+}_{N}(\omega,T)$ for doping values of $\delta=0.3,0.2,0.15,0.1$.}
\label{current_correlation}
\end{figure}
The imaginary part of current-current correlation function, $\Im \chi^{+}_{N}(\omega)$, and the real part of the optical conductivity, $\sigma(\omega)$ are defined as:
\begin{equation}
\frac{\sigma(\omega)}{\pi\sigma_0}=\frac{(1-e^{-\beta\omega})}{\omega}\Im \chi^{+}_{N}(\omega)
\end{equation}

In this expression, $\sigma_0$ can be taken to be of order $\sigma_{0} = \frac{a^{2-d}e^2}{\hbar}$, with the lattice spacing $a$ (corresponding in a quasi two-dimensional system to a sheet resistance of one quantum per plaquette). In Fig $\ref{current_correlation}$, we present contour plots of imaginary part of current-current correlation function $\Im \chi^{+}_{N}(\omega)$ for various doping values. The spectra display the following features: (a) For $T<\omega$, $\Im \chi^{+}_{N}(\omega)$ varies with $\omega$ up to a certain $\omega$ and is constant afterwards. (b) For small $\omega$ and $T>\omega$, $\Im \chi^{+}_{N}(\omega)$ is a constant in temperature. This region of constant $\Im \chi^{+}_{N}(\omega)$ reduces as doping is reduced. (c) For high $T$ and small $\omega$, we see another region of constant $\Im \chi^{+}_{N}(\omega)$. We conclude that resistivity when $\Im \chi^{+}_{N}(\omega)$ is independent of temperature is linear in $T$. As mentioned above, the region of linear $T$ is large for large doping and starts reducing upon decreasing doping. Our approach finds two regimes of linearity in resistivity, with different slopes, one at intermediate and another at high temperatures. It should also be noted that the Planckian constant is the inverse of $\Im \chi^{+}_{N}(\omega)$, and it should be the same for all doping an from our results, it is not exactly one, but is close to it for higher doping but not for lower doping.

\section{Self Energy and DC Resistivity:}
\label{section4}

 In this Section, we discuss the electron self energy and the {\bf intrinsic} dc electrical resistivity of the infinitely correlated metal, which is intimately linked with the electron self energy. For instance, the imaginary part of the self-energy, $\Im\Sigma(\omega,T)$, provides insight into the lifetime and coherence of quasiparticles, which are crucial for determining how electrons propagate through a material, and so its resistivity. 
 
We also note here (subsection A) that analytically (see Appendix \ref{AppendixD}) the imaginary part of the self-energy, $\Im \Sigma(\omega,T)$, adheres for very low $\omega$ and $T$ to the Fermi liquid description, scaling as $(\omega^2 + \pi^2 T^2)$. This is also seen from our self consistent result for $\Im \Sigma(0,T)$. This insight sets the stage for a deeper analysis of how single particle properties change with doping, for both  positive and negative excitation energies (particle like and hole like) and at different temperatures. Some of the results are exhibited in Fig \ref{Selfenergy_spectralfunc} parts (a) to (d), and insets therein. For example, we  show that in the Fermi liquid (very low temperature) regime the quasiparticle residue $Z$ (typically of order 0.1 to 0.2) increases roughly linearly with increasing hole doping. We also plot the local single particle spectral density and see the quasiparticle like low excitation energy peak in it disappearing as temperature increases and the electron system becomes an incoherent liquid of Fermi like excitations. 

We next discuss (subsection B)  dc resistivity using the well known large $d$ form for the result \cite{DMFTREVIEW} which neglects vertex corrections. At low temperatures, it is seen to have the classic Fermi liquid form, going as $T^{2} $(in correspondence with the result above for the same region, namely that $\Im \Sigma(\omega,T)$ goes as  $(\omega^2 + \pi^2 T^2)$. It transitions via a long crossover region straddling both the incoherent quantum regime and the `classical' metal regime (Section \ref{section3}) into linear resistivity behaviour. We also see no signs of resistivity saturation; the resistivity continues to rise linearly with the same slope, beyond the Mott-Ioffe Regel (MIR) quantum limit. 

We believe that the defining characteristics of the extremely strongly correlated metal, mentioned above, are due to the influence of local bosonic charge and spin fluctuations which are strongly coupled to electrons. They determine the electron self energy (Section \ref{section 2}) and have a sizeable, nearly constant, strength over a large frequency region at most temperatures.

\subsection{Scattering Rate and Local Bosonic Correlation Functions}

\begin{figure}
\centering
\includegraphics[width=0.85\textwidth]{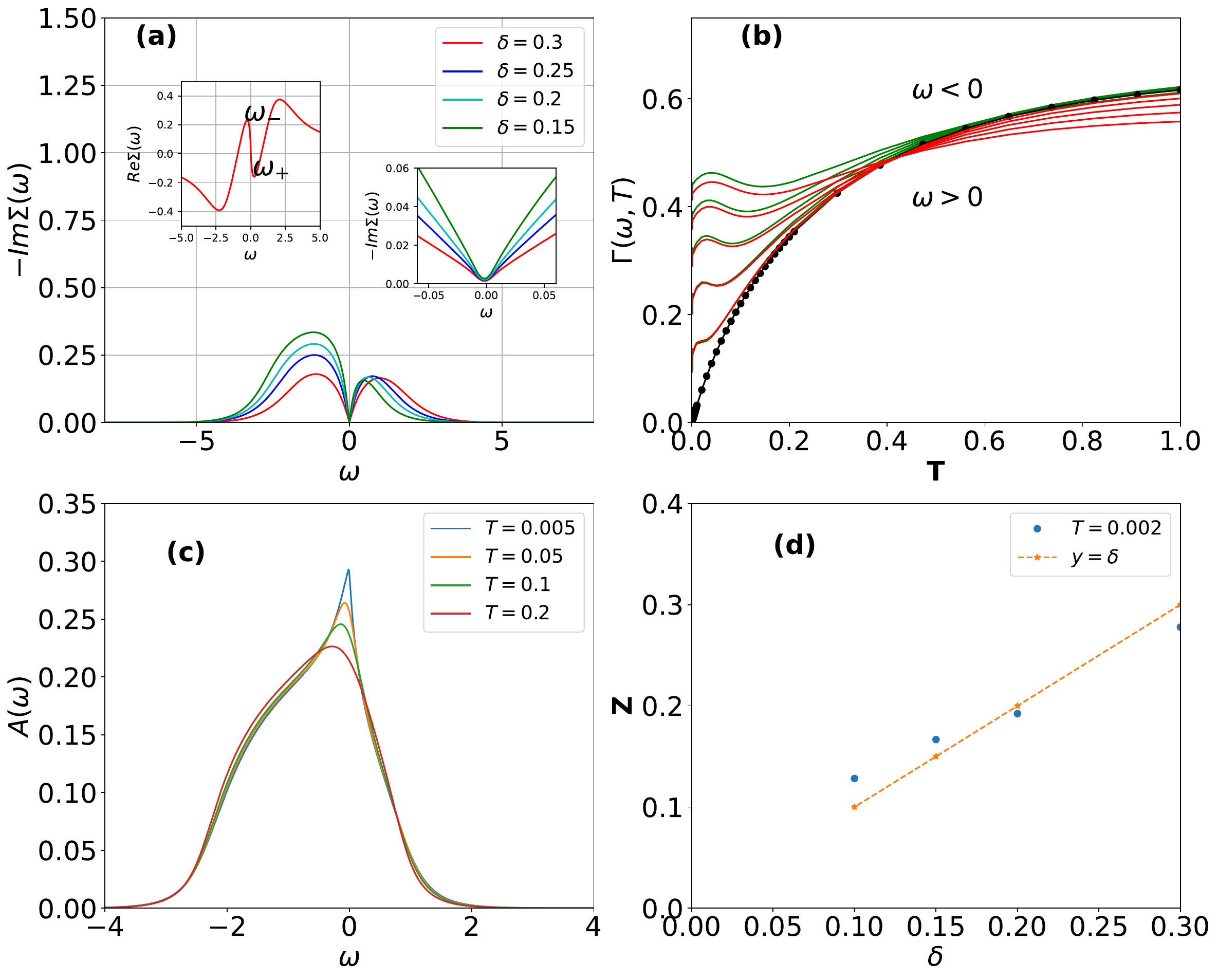}
\caption{(a) $-\Im \Sigma(\omega)$ at $T=0.005$ for various doping levels. The left inset shows $\Re \Sigma(\omega)$, 
linear in $\omega$ for $\omega_{-}<\omega<\omega_{+}$. The right inset is $-\Im \Sigma(\omega)$ near $\omega=0$ showing $\omega^2$ behavior 
. (b) $-\Im\Sigma(\omega_c,T)$ versus $T$ at $\delta=0.2$ for $\omega_c = -0.5, -0.4, ..., -0.1$ (green), $\omega_c = 0.0$ (thick black), and $\omega_c = 0.1, 0.2, ..., 0.5$ (red). (c) Spectral function at doping $\delta=0.2$ at various temperatures. (d) Variation of Quasiparticle weight with doping at $T= 0.002$.}
\label{Selfenergy_spectralfunc}
\end{figure}

In Figure \ref{Selfenergy_spectralfunc}(a), start going into the details of
$-\Im \Sigma(\omega)$, examining how it varies with frequency ($\omega$) across diverse doping levels at a notably low temperature of $T = 0.005$. This analysis uncovers a fascinating transition in behavior around $\omega=0$, where the pattern evolves from square-like to linear. Such a transformation underscores the pivotal role of local charge dynamics in modulating electron scattering processes. Interestingly,  when comparing the degree of particle-hole asymmetry in our findings with those obtained from Dynamical Mean Field Theory (DMFT) and Shastry's work, ours exhibit a lesser asymmetry.
Furthermore, within the same graphical representation, we shift our focus to the real component of the self-energy, $ \Re \Sigma(\omega)$. Here, we identify a linear section extending between two critical points, $\omega_{-}$ and $\omega_{+}$, with the latter's value notably adjusting in response to variation in doping levels.

Venturing into Figure \ref{Selfenergy_spectralfunc}(b), our exploration extends to the imaginary component of self-energy at a specific frequency, evaluated as a function of temperature. This provides insight into the behavior of the scattering rate under finite frequencies. Notably, for hole-like excitations (where $\omega<0$), the scattering rate consistently exceeds that of electron-like excitations (where $\omega>0$). As temperature increases, the relationship between $ \Im \Sigma$ and $T$ for various positive frequencies unveils a crossing threshold. Beyond it, the scattering rate inversely correlates with frequency, implying that at higher temperatures, low-energy, electron-like excitations with finite positive $\omega$ values enjoy longer lifespans compared to those precisely at $\omega=0$. While the overall trends are the same as DMFT predictions, some of our findings differ; for example, the $\Gamma(\omega,T)$ crossing is at much larger values of $ \omega$ and $T$.
 
In Figure \ref{Selfenergy_spectralfunc}(c), the progression of the spectral function with varying temperatures is presented, focusing specifically on a doping level of $\delta=0.2$. This visualization allows us to observe how temperature influences the spectral features, revealing important insights into the decoherence effects on the electronic structure at this particular level of doping. On the other hand, Figure \ref{Selfenergy_spectralfunc}(d) is dedicated to illustrating the behavior of the quasi-particle residue across a range of doping levels at a particular low temperature (T= 0.002) where the system is a Fermi liquid. This aspect of the study sheds light on the correlation between doping concentration and the quasi-particle strength, elucidating how the electronic properties of the system evolve with changes in doping.

\subsection{DC Resistivity and the Influence of Local Bosonic Correlation Functions}

\begin{figure}
\centering
\subfloat[]{\includegraphics[width=0.42\textwidth]{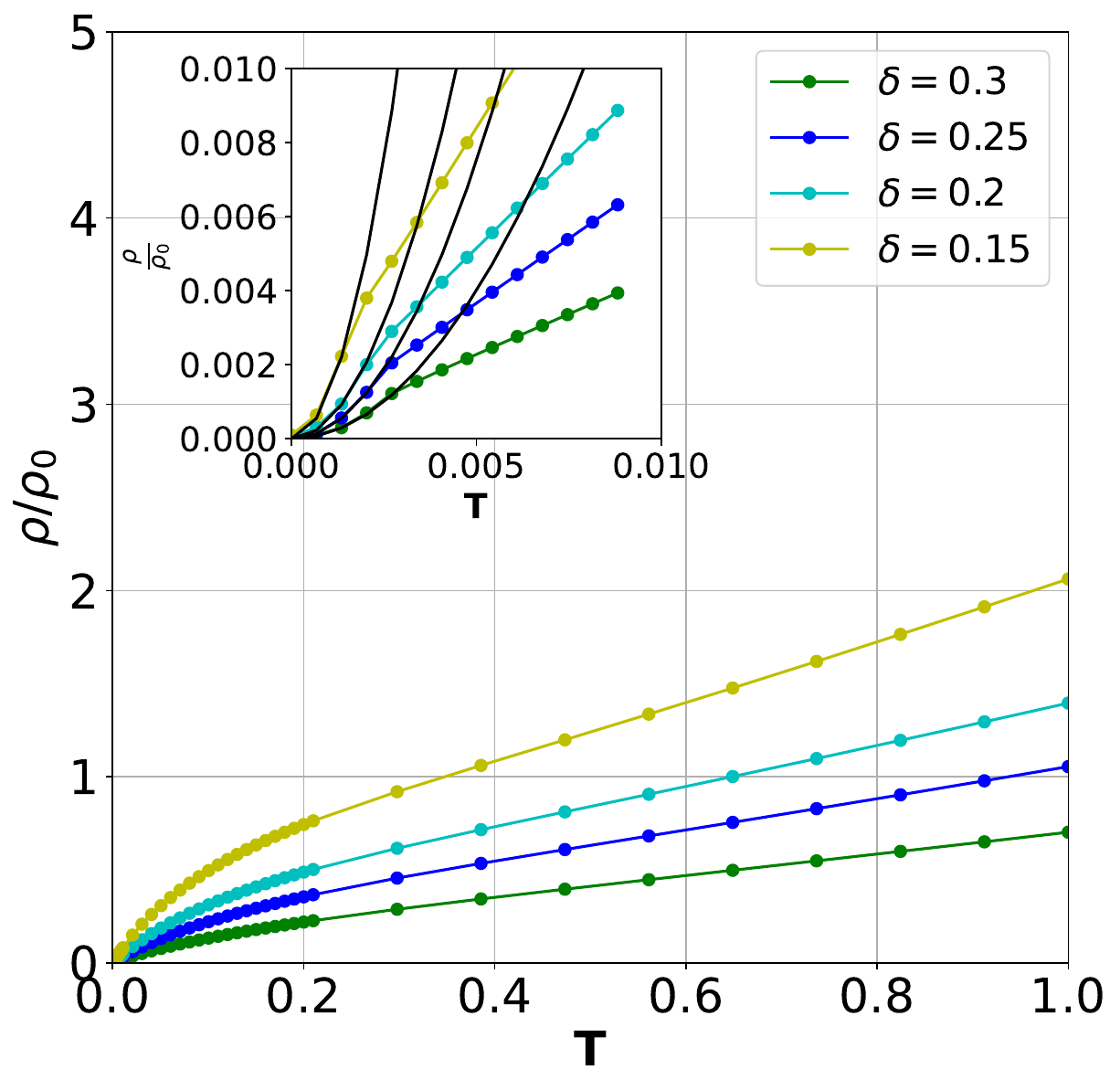}}
\quad
\subfloat[]{\includegraphics[width=0.45\textwidth]{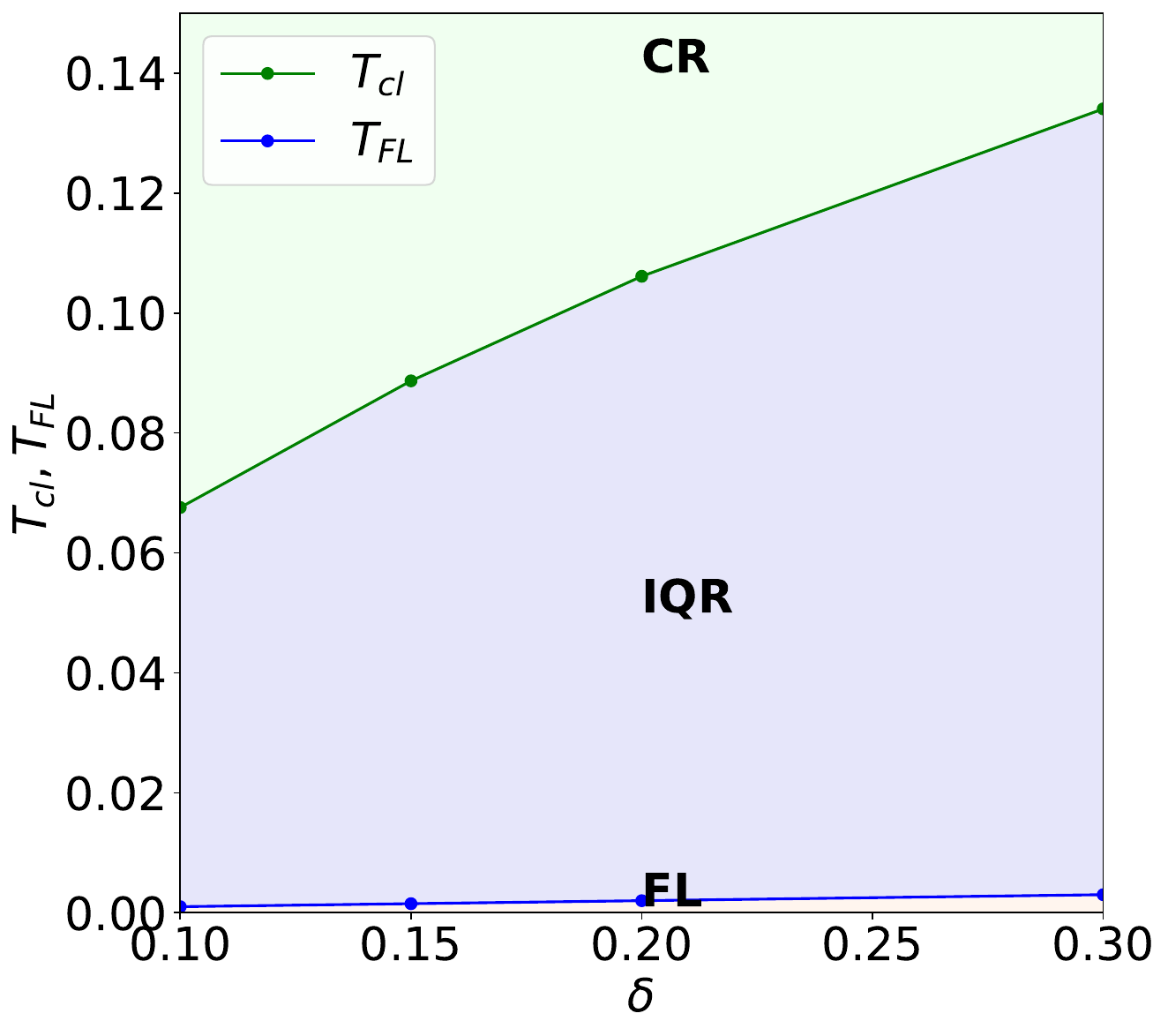}}
\caption{(a) Temperature dependence of resistivity for several doping levels $\delta$ in the unit of $\rho_0(=1/\sigma_0)$. The inset shows the low-temperature resistivity vs. $T$, revealing the $T^2$ behavior with the black line representing the parabolic fit. (b) Different temperature regimes: FL (yellow) for $T<T_{FL}$, incoherent quantum regime (IQR) for $T_{FL}<T<T_{cl}$, and $T>T_{cl}$ a classical regime (CR).}
\label{scatrate}
\end{figure}

In  Figure \ref{scatrate}(a), we exhibit the relationship between resistivity ($\rho$) and temperature ($T$) for various levels of doping. At the lowest temperatures, the system demonstrates typical Fermi-liquid behavior, characterized by a quadratic increase in resistivity with temperature. This trend is clearly depicted in the inset of Figure \ref{scatrate}(a), in which the low temperature region is shown, enlarged,
illustrating the coherent interactions among particles. These interactions are mediated by local charge bosons (charge excitations), which play a critical role in the coherent Fermi liquid like resistivity behaviour of the the system at low temperatures as well. 
As temperature increases, crossing into the Incoherent Quantum Regime (IQR), we observe a linear rise in resistivity. This change signals a crossover from coherent to incoherent or chaotic behavior, despite the continued influence of local charge excitations on the system's dynamics.

Analyzing the resistivity outcomes more closely, we distinguish three separate temperature domains as outlined in the data: $0<T<T_{FL}$, where $T_{FL}$ signifies the temperature boundary above which Fermi-liquid behavior is no longer observed; $T_{FL}<T<T_{IQR}$, marking the range within the Incoherent Quantum Regime characterized by a linear increase in resistivity; and $T>T_{cl}$, representing temperatures beyond which the system exhibits incoherent behavior (see Section 2). These domains are graphically represented as functions of doping in  Figure \ref{scatrate}(b), where the temperature thresholds $T_{FL}$ and $T_{cl}$ are plotted against doping levels. It's observed that $T_{FL}$ changes linearly with doping, indicating a direct correlation between doping levels and the Fermi-liquid to non-Fermi-liquid crossover temperature. In contrast, the variation of $T_{cl}$ with doping does not follow a simple linear pattern, underscoring the complex interplay between doping and the material's crossover to incoherent electronic states.

This extensive exploration into self-energy, electron scattering, and resistivity, all through the lens of local charge excitations, uncovers the complex dynamics in extremely correlated electron systems. Its impact is seen to vary in a characteristic way with
temperature, frequency, and doping on electron behavior, crossing over from coherent to incoherent or chaotic states, and presents a thorough framework for comprehending the diverse phenomena observed in these complex materials, centering around the role of local charge excitations.


\section{Summary, Limitations and Future Directions}
\label{section 5}

In our research, we explore metals with infinitely strong local electron repulsion, known as extremely correlated Fermi liquids (ECFL), employing for it the single orbital Hubbard model at $U=\infty$. This model, and  the non-interacting model (which is at the other limit of $U$, namely with $U=0$) are both characterized by a single parameter, the chemical potential which governs the average electron count in the system. The $U=0$ limit,
 foundational to the Drude model, serves as the baseline for developing perturbative theories for electron systems with interactions ($U\neq 0$), simplifying interaction effects into a concise set of Landau parameters.

We propose a new, approximate, self-consistent theory using the nonperturbative equation of motion technique for lattice quantum fields, resulting in an equation for the Dysonian self-energy $\Sigma$ of the single-particle Green's function $G$. We find that $\Sigma$ is roughly the convolution of local bosonic degrees of freedom, namely of local charge and spin fluctuation correlators $D_{N}$ and $D_{S}$, with $G$, an approach proven exact in the $d=\infty$ limit. Subsequently, we derive equations for the correlators $D_{N}$ and $D_{S}$, disregarding vertex corrections. We find that these are electron hole fluctuations, and iterate to a self-consistent solution. This is the mechanism for electron dynamics in strongly correlated systems.  
 ' 
We find that in the strong correlation limit, spatially local diffusive, quantum, bosonic fluctuations made up of electron hole excitations with their strong coupling to electrons, determine the characteristic dynamics. This dynamics is also temporally nearly local; the diffusive fluctuation spectrum (`noise') extends with significant and nearly constant strength over a wide frequency range for all (except rather low ) temperatures. We reach this conclusion by making several simplifying approximations which enable us to clearly focus on these fluctuations: we consider  only the simplest `scattering correction' term involving this fluctuation  for the electron self energy; we use the large $d$ approximation so that the bosonic correlator of relevance is atomic site local and is separated out; we use a self consistent approach to evaluate both the electron propagator and the bosonic correlator to emphasize their strong interrelation. We also use throughout a real time description to emphasize the physical domain of their operation, obviating the need for analytical continuation from imaginary time (and in frequency space, from discrete imaginary Matsubara frequencies). Further, since there is no restoring force or characteristic frequency of the bosonic mode, the characteristic temperature can be low, determined solely by when the long time or low energy quantum coherence due to these diffusive local fluctuations becomes ineffective.

This seems quite different from a recent approach by Subir Sachdev and coworkers (see for example some recent papers , e.g. refs.\cite{Patel1}, \cite{Sachdev1} \cite{Patel2}) who basically propose and develop a model for electrons coupled to zero energy (or quantum critical) phonons; the basic Yukawa electron phonon coupling has static disorder, as does the electron system. They do a very sophisticated weak coupling  analysis of the resulting  many body system, and obtain some of the above broad results, namely a crossover from Fermi liquid to incoherent metal behaviour, and linear resitivity with the potential to be universal. There are superficial similarities in that the many body theory leads to overdamped, localized bosonic modes, for example. However, in our case, there is no static disorder and no static localization; the observed properties emerge as consequences of quantum dynamics in the strongly correlated clean system . There are no zero frequency or quantum critical bosons; the bosonic excitation is self generated. There is no extreme phonon drag regime. Ours is a materials based strong coupling approach, albeit captured in an oversimplified one orbital Hubbard model. It is not impossible, though, that because of a kind of duality, there may be some mapping between the strong coupling and the weak coupling approaches which makes sense of the similarity in the most significant outcomes.

Our findings highlight two distinct characteristics of strongly correlated systems: the crossover from coherent to incoherent Fermi liquid behavior as temperature increases, and the emergence of linear resistivity in the strange metal phase, both predominantly due to the coupling of electrons with local number fluctuations. Earlier results, supported by numerical analyses (e.g.single site  DMFT (\cite{Antoine}), DCA (\cite{Tremblay})), emphasize the crucial influence of onsite, incoherent charge fluctuations ($D_{N}$ and $D_{S}$) on the electronic properties.The qualitative outcomes of these, and of the line of work by Shastry and collaborators (\cite{Shastry1}, \cite{Shastry2}, \cite{Shastry3}) are the same as ours.
 
We observe a potential universality in ECFL properties, such as linear resistivity, resulting from localized, incoherent charge noise driven by hole movement, highlighted by the straightforward dependence of physical properties on hole density (Sections \ref{section3} - \ref{section4}). At temperatures beyond the quantum scale of local fluctuations, the noise spectrum becomes `white', reflecting classical thermal noise driven by hole motion (see Fig. 4 for scale comparison). This suggests a universal interaction mechanism between local fluctuations and electrons, leading to fluctuations in local electron numbers due to electron dynamics.

However, our results may not directly correlate with observations from specific systems; one fact is the higher crossover temperature $T_{cl}^{*}$ compared to empirical findings for the onset of the strange metal regime. This discrepancy could be attributed to our excluding the orthogonality catastrophe effect ( this was pointed out first by Anderson (\cite{Anderson}) who included it approximately in a Gutzwiller projected fermion approach. Additionally, since all strongly correlated systems have a large but finite $U$, there is quite likely to be a new low energy scale related to it, specifically the intersite spin coupling scale $J_{ij}$, which sets a new small temperature scale, of relative order $(t/U)$.

Our ongoing research aim is to investigate the role of local bosonic excitations in systems with large, finite, $U$ values, using hybridization-expansion continuous-time quantum Monte Carlo (CTQMC) and the local moment approach within dynamical mean-field theory. The latter, albeit approximate, has the advantage of yielding real frequency spectra and self-energies at zero and finite temperatures, and hence will be complementary to CTQMC, which yields Matsubara frequency quantities and requires analytic continuation. We also aim to develop a model for the leading $(1/U)$ phenomena in the spirit of quantum mechanical perturbation theory, with the $U=\infty$ results of this paper as the unperturbed system input.

\section*{Acknowledgements}

Our sincere thanks go to M.S. Laad, H.R. Krishnamurthy, Mohit Randeria, Nandini Trivedi, Subroto Mukherjee, and Arindam Ghosh for  insightful and invaluable discussions. TVR is grateful for the support received from the DST as a Year of Science Chair Professor during the initial phase of this work. SRH appreciates the hospitality and access to facilities provided by JNCASR during visits.

\bibliographystyle{apsrev4-2}
\bibliography{draft}

\appendix

\section{\texorpdfstring{$X$}{Xop} Operators }
\label{Appendix A}

  For a lattice system with one orbital per site, a general state can be described completely 
  in terms of the orbital states at a site $i$. This set consists of states $\ket{0}$, $ \ket{\sigma}$, $\ket{\bar{\sigma}}$ and $\ket{2}$; namely those with no electron, one electron with spin $\sigma (\uparrow)$ or $\bar{\sigma}(\downarrow)$, and two electrons($\uparrow \downarrow$). The $X$ operators introduced by Hubbard \cite{Hubbard2} are all the matrix elements in this Hilbert space; e.g. $X_{i}^{\sigma 0}$ is $\ketbra{\sigma}{0}$ for states at site $i$. They are local Fermi like or Bose like field operators (not canonical  Fermi or Bose operators), depending on whether they describe change in local electron number by unity (odd numbers in general) or by zero (even numbers in general) ( see the book by Ovchinnikov and Val'kov \cite{Ovchinnikov} is on the $X$ operators and its application in condensed matter physics). Commutators/anticommutators of $X$ operators at different sites vanish, while for the the same site, they do not. These results are uniquely determined by the definition of $X$ operators. The results of on site commutation/anticommutation are not $c$ numbers as for canonical fermions and bosons, but are $X$ operators.

The $X$ operators obey the commutation relation 
\begin{align}
    \commutator {X_{i}^{\alpha\beta}} {X_{j}^{\gamma\delta}}_{\pm} = \qty( X_{i}^{\alpha\delta}\delta_{\beta\gamma} \pm X_{i}^{\gamma\beta}\delta_{\delta\alpha}) \delta_{ij} \mbox{,}
\label{eqn1A}
\end{align}
since at a given site one has
\begin{align}
X_{i}^{\alpha\beta}X_{i}^{\gamma\delta} =\delta_{\beta\gamma}X_{i}^{\alpha\delta}\mbox{.}
\label{eqn2A}
\end{align}
The basic commutator involving $X$ arises from the Heisenberg equation of motion for the $X$-operator. It is 
\begin{align}
   i\partial_{t}X_{i}^{\alpha\beta}=i\dot{X}_{i}^{\alpha\beta}= \commutator {X_{i}^{\alpha\beta}} {H}_{-} &=-\mu\sum_\sigma \qty (X_{i}^{0 \sigma}\delta_\beta\sigma -X_{i}^{\sigma\beta} \delta_{\sigma\alpha})\nonumber\\
    &+\sum_{jm}t_{jm} \qty (\commutator {X_{i}^{\alpha\beta}} {X_{j}^{0 \sigma}}_{\pm} X_{m}^{\sigma 0} \pm X_{j}^{0 \sigma}  \commutator {X_{m}^{\sigma 0}} {X_{i}^{\alpha\beta}}_{\pm}) \mbox{.}
\label{eqn3A}
\end{align}
where the upper and lower set of signs are for bosonic and fermionic operators respectively.\footnote{We use the (nearly standard) convention that $[A,B]_{\pm}$ is an anticommutator for the + sign and the commutator for the - sign.}

The case of $\alpha = 0$ and $\beta = \sigma$, namely the fermionic operator ${X}_{i}^{0\sigma}$ is relevant for the equation of motion of the single particle Green\textquotesingle s function. We have 
\begin{align}
\comm{X_{i}^{0\sigma}}{X_{j}^{\sigma^{\prime} 0}}_{+} = \delta_{ij}(\delta_{\sigma\sigma^\prime} X_{i}^{00} + X_{i}^{\sigma' \sigma}) = \delta_{ij} B_{i}^{\sigma\sigma^\prime}
\label{eqn4A}
\end{align}

where $B_{i}^{\sigma\sigma^\prime}$ is a bosonic operator centred at $i$. It is a charge fluctuation operator for $\sigma = \sigma^\prime$  and a spin fluctuation operator for $\sigma \neq \sigma^\prime$. Using the compact notation ref.\cite{Shastry1}, this can be written as 
\begin{align}
    \comm{X_{i}^{0\sigma}}{X_{j}^{\sigma^{\prime} 0}}_{+} = \delta_{ij}(\delta_{\sigma\sigma^\prime} I -\sigma\sigma^\prime X_{i}^{{\bar\sigma}{\bar\sigma^\prime}})
\end{align}

 \begin{align}
i\dot{X}_{i}^{0\sigma}(t) = -\mu X_{i}^{0\sigma}(t) + \sum_{m\sigma^\prime} t_{mi} B_{i}^{\sigma\sigma^\prime}(t)X_{m}^{0\sigma^\prime}(t) \mbox{,}
\label{eqn5A}
 \end{align} 
 and
\begin{align}
\comm{B_{i}^{\sigma_{1}\sigma_{2}}}{X_{j}^{0\sigma_3}}_{-}=-\sigma_{1}\sigma_{2} \comm{X_{i}^{{\bar\sigma_1}{\bar\sigma_2}}}{X_{j}^{0\sigma_3}}_{-}=
-\sigma_{1}\sigma_{2}\delta_{ij} X_{i}^{0\bar\sigma_2}\delta_{{\bar\sigma_1}\sigma_{3}}.
\label{eqn6A}
\end{align} 

In equation (\ref{eqn5A}),we notice that there is a novel, local spin flip term due to hopping (last term on the right) present only because of correlation. This involves a spin flip at say site $i$ and a number change (of the spin flipped electron)  at site $j$ connected with it via hopping. In the following, we assume (as is common) that $t_{im} = t_{mi}$. 
  
The equation of motion for bosonic operators is illustrated with the example of $\dot{X}_{i}^{\bar{\sigma} \sigma}$ for which 
  $\alpha = \bar{\sigma}$, $\beta = \sigma$. We have
\begin{align}
  -\dot{X}_{i}^{\bar{\sigma}\sigma} = \sum_{m} t_{im}\qty (X_{m}^{0 \sigma}X_{i}^{\bar{\sigma} 0}-X_{i}^{0 \sigma}X_{m}^{\bar{\sigma} 0}) \mbox{.}
\label{eqn7A}
 \end{align}
 
For the extremely strongly correlated Fermi liquid (ECFL) where $(U/t) \rightarrow \infty$ the doubly occupancy state $\ket{2}$ can be neglected since it has an infinitely high energy. In this limit, the relations satisfied by the Hubbard operators can be written as 
\begin{align}
X_{i}^{00}+\sum_{\sigma}X_{i}^{\sigma\sigma} =I\mbox{,}\hspace{0.1cm}
X_{i}^{\sigma\sigma}+X_{i}^{\bar{\sigma}\bar{\sigma}} = {N}_{i}\mbox{,}\hspace{0.1cm}
X_{i}^{\sigma\sigma}-X_{i}^{\bar{\sigma}\bar{\sigma}}=2S_{i}^z \mbox{,}
\label{eqn8A}
\end{align}
\begin{align}
X_{i}^{00}+\sum_{\sigma}\sigma X_{i}^{\sigma\sigma}& = I-\frac{N_{i}}{2}+\sigma S_{i}^{z} \mbox{.}
\label{eqn9A}
\end{align}
where $I$ is the identity operator, and $N_i$ and $S_i^z$ are respectively the number and the $z$ component of spin operators at site $i$. Since the system is homogeneous, the thermodynamic average at any site $i$ is independent of $i$. We define $n$ and $m$ as the average number and the average $z$ component of the magnetization, namely $n= \expval{N_i}$ and $m=\expval{S^{z}_i}$. We assume that the system is paramagnetic, so that $m =0$ and $\expval{X^{\bar{\sigma}\bar{\sigma}}}  = \expval{X^{\sigma\sigma}} = n/2$, and that it is spin isotropic. A commonly occurring quantity is
\begin{align}   
  \expval{B_{i}^{\sigma\sigma}} =
 \expval{I-\frac{N_{i}}{2}+S_{i}^{z}}
  = \qty(1 - \frac{n}{2}) = \qty(\frac{1 + \delta}{2})= Q \mbox{.}
\label{eqn10A}
\end{align}

\section{Equation of motion for \texorpdfstring{$D^{+}_{N}$}{Dplus} and \texorpdfstring{$D^{+}_{S}$}{Dminus}}
\label{Appendix B}

$D^{R}$ in terms of the $N$ and $S_{z}$ operators can be derived using $B^{\sigma \sigma} = X^{00} + X^{\sigma \sigma} = 1 - \frac{N}{2} + S_{z},~ B^{\sigma \Bar{\sigma}} = X^{\sigma \Bar{\sigma}} = S^{+}$ for B-operator.
\begin{align}
    D^{R}(t,t^\prime) & = \expval{\expval{B^{\sigma \sigma}(t)|B^{\sigma \sigma}(t')}} + \expval{\expval{B^{\sigma \Bar{\sigma}}(t)|B^{\Bar{\sigma} \sigma}(t')}}
    \\  D^{R}(t,t^\prime) & = \frac{1}{4}\expval{\expval{N(t)|N(t')}} + \expval{\expval{S_{z}(t)|S_{z}(t')}} +  \expval{\expval{S^{+}(t)|S^{-}(t')}} - \frac{1}{2} \expval{\expval{N(t)|S_{z}(t')}} - \frac{1}{2} \expval{\expval{S_{z}(t)|N(t')}}
    \\D^{R}(t,t^{\prime}) &= \frac{1}{4} \expval{\expval{ N(t) | N(t^\prime)}} + \frac{3}{2} \expval{\expval{S^{+}(t) | S^{-}(t^\prime)}} 
     -\frac{1}{2}\expval{\expval{ N(t) | S^{z}(t^\prime)}}  - \frac{1}{2}  \expval{\expval{ S^{z}(t) | N(t^\prime)}}
     \end{align}
Since $\expval{\expval{S^{+} | S^{-}}} = \expval{\expval{S_{x} | S_{x}}}  + \expval{\expval{S_{y} | S_{y}}} = 2 \expval{\expval{S_{z} | S_{z}}}$ in isotropic phase.
The third and fourth terms would vanish in the paramagnetic phase.

\begin{align}
    D^{R}(t,t^{\prime}) &= \frac{1}{4} \expval{\expval{N(t) | N(t^\prime)}} + \frac{3}{2} \expval{\expval{ S^{+}(t) | S^{-}(t^\prime)}} 
     \end{align}

Since,
\begin{equation}
    \begin{aligned}
        D^{+}_{N}(t-t') = - i \theta (t - t') \expval {N(t) N(t')}
        \\ D^{+}_{S}(t-t') = - i \theta (t - t') \expval {S^{+}(t) S^{-}(t')}
    \end{aligned}
\end{equation}

To develop the equation of motion for $D^{+}_{N}$ we start by differentiating it with respect to $t$ 

\begin{equation}
\begin{aligned}
    i \partial_{t} D^{+}_{N}(t-t') & = \delta (t-t') \expval{N N} - i \theta(t-t') \expval{i\dot{N}(t) N(t')}
    \\ \implies i \partial_{t} D^{+}_{N}(t-t') & = \delta (t-t') n - i \theta(t-t') \expval{i\dot{N}(t) N(t')}
    \\ \implies i \partial_{t} D^{+}_{N} (t-t') & = \delta (t-t') n + \Tilde{D}^{+}_{N} (t-t')
\end{aligned}
\label{D3}
\end{equation}

where $n$ is the number density and 

\begin{align}
    \Tilde{D}^{+}_{N}(t-t') = - i \theta(t-t') \expval{i\dot{N}(t) N(t')}
\end{align}

Now, we develop an equation of motion for $\Tilde{D}^{+}_{N}$ by differentiating it with respect to $t'$

\begin{equation}
    \begin{aligned}
        i \partial_{t'} \Tilde{D}^{+}_{N} (t-t') = - \delta (t-t') \expval{i \Dot{N}(t) N(t)} + i \theta (t-t') \expval {\Dot{N}(t) \Dot{N}(t')}
    \end{aligned}
\end{equation}

First term in the above expression is zero which can be easily seen from the fact that the number operator acting on number basis (in which we are taking the trace) will give us the same state so $\expval{\Dot{N}N} = \expval{\Dot{N}}$ which is $0$ in equilibrium , so we get

\begin{equation}
\begin{aligned}
     i \partial_{t'} \Tilde{D}^{+}_{N} (t-t') = i \theta (t-t') \expval {\Dot{N}(t) \Dot{N}(t')}
     \\ i \partial_{t'} \Tilde{D}^{+}_{N} (t-t') = - \chi^{+}_{\scriptscriptstyle JJ} (t-t')
\end{aligned}    
\label{D6}
\end{equation}

where we have defined 

\begin{align}
    \chi^{+}_{\scriptscriptstyle JJ} (t-t') = -i \theta (t-t') \expval {\Dot{N}(t) \Dot{N}(t')}
\end{align}

We define the Fourier transform as (in the same manner for all the terms)

\begin{align}
    D^{+}_{N}(t-t') = \frac{1}{2 \pi} \int \dd (t-t') e^{-i \omega (t-t')} D^{+}_{N}(\omega)
\end{align}

Fourier transforming and combining equations (\ref{D3}) and (\ref{D6}) we get

\begin{align}
    D^{+}_{N}(\omega) = \frac{1}{\qty[ \frac{\omega}{n} - \frac{\chi^{+}_{\scriptscriptstyle JJ}(\omega)}{n^{2}}]}
\end{align}

Similarly, for $D^{+}_{S}$ 

\begin{equation}
\begin{aligned}
     i \partial_{t} D^{+}_{S}(t-t') & = \delta (t-t') \expval{S^{+} S^{-}} + \Tilde{D}^{+}_{S}(t-t')
     \\ \expval{S^{+} S^{-}} & = 2 \expval{S^{z} S^{z}} = \frac{\expval{X^{\sigma \sigma} + X^{\bar{\sigma} \bar{\sigma}}}}{2}  = \frac{n}{2}
     \\ \implies i \partial_{t} D^{+}_{S}(t-t') & = \delta (t-t') \frac{n}{2} + \Tilde{D}^{+}_{S}(t-t')
\end{aligned}
\label{D10}
\end{equation}

where 

\begin{align}
    \Tilde{D}^{+}_{S}(t-t') = -i \theta(t-t') \expval{i \Dot{S}^{+}(t) S^{-}(t')}
\end{align}
\begin{align}
i \partial_{t'} \Tilde{D}^{+}_{S}(t-t') = -\delta(t-t') \expval{i \Dot{S}^{+} S^{-}}  + i \theta(t-t') \expval{\Dot{S}^{+}(t) \Dot{S}^{-}(t')}   
\end{align}

To determine the $\expval{i \Dot{S}^{+} S^{-}}$ we go back to the definition $S^{+} = X^{\sigma \bar{\sigma}}$ and $S^{-} = X^{\bar{\sigma} \sigma}$. using Heisenberg equation of motion, we find

\begin{align}
    i \Dot{X}^{\sigma \bar{\sigma}} = \sum_{j} t_{ij} \qty (X_{i}^{\sigma 0} X_{j}^{0 \bar{\sigma}} - X_{i}^{0 \bar{\sigma}} X_{j}^{\sigma 0})
\end{align}

using this relation, we can find $i \Dot{S}^{+}S^{-} = i \Dot{X}^{\sigma \bar{\sigma}} X_{i}^{\bar{\sigma} \sigma} $ to be 

\begin{align}
    i \Dot{S}^{+}S^{-} = \sum_{j} t_{ij} \qty(X_{i}^{\sigma 0} X_{j}^{0 \sigma} - X_{j}^{\sigma 0} X_{i}^{0 \sigma})
\end{align}

rhs of the above equation is same as the current operator, hence $\expval{i \Dot{S}^{+} S^{-}} = 0$. so, we have 

\begin{equation}
\begin{aligned}
    i \partial_{t'} \Tilde{D}^{+}_{S}(t-t') & =  i \theta(t-t') \expval{\Dot{S}^{+}(t) \Dot{S}^{-}(t')}
    \\ i \partial_{t'} \Tilde{D}^{+}_{S}(t-t') &= - \chi^{+}_{\scriptscriptstyle J_{s}J_{s}}(t-t')
    \\ \chi^{+}_{\scriptscriptstyle J_{s}J_{s}} (t-t') & = - i \theta(t-t') \expval{\Dot{S}^{+}(t) \Dot{S}^{-}(t')}
\end{aligned}    
\label{D15}
\end{equation}

Fourier transforming and combining equations (\ref{D10}) and (\ref{D15}) , we get

\begin{align}
    D^{+}_{S}(\omega) = \frac{1}{\qty[\frac{\omega}{n/2} - \frac{\chi^{+}_{\scriptscriptstyle J_{s}J_{s}}(\omega)}{n^{2}/4}] }
\end{align}

\section{Current current correlation function for spin and charge}
\label{Appendix C}

In imaginary time, we can write the time ordered current-current correlation functions for charge and spin as 
\begin{align}
    \chi_{N}(\tau , \tau') = - \expval{T_{\tau} J_{c}(\tau) J_{c}(\tau')} \mbox{,} \quad
    \chi_{S}(\tau , \tau') = - \expval{T_{\tau} J_{s}(\tau) J_{s}(\tau')}
    \\  J_{c} = \frac{1}{N}\sum_{k,\sigma}v_{k}X_{k}^{0\sigma}X_{k}^{\sigma0} \mbox{,} \quad J_{s} = \frac{1}{N}\sum_{k}v_{k}X_{k}^{0 \sigma}X_{k}^{\bar{\sigma}0} 
\end{align}
$T_{\tau}$ is the time ordering operator. The contribution to bubble diagram in imaginary frequency is 
\begin{align}
    \chi_{N}(i \nu_{n}) = \frac{1}{N}\sum_{\sigma,k,m} v_{k}^{2} G^{\sigma \sigma} (k,i \omega_{m}) G^{\sigma \sigma}(k, i \nu_{n} + i \omega_{m}) 
    \\ \chi_{S}(i \nu_{n}) = \frac{1}{N}\sum_{k,m} v_{k}^{2} G^{\sigma \sigma} (k,i \omega_{m}) G^{\bar{\sigma} \bar{\sigma}}(k, i \nu_{n} + i \omega_{m}) 
\end{align}
$i \omega_{n} = \frac{(2n+1)\pi}{\beta}$, $i \nu_{n} = \frac{2n\pi}{\beta}$ and $\beta$ is inverse temperature. Since in the paramagnetic phase $G^{\sigma \sigma} = G^{\bar{\sigma} \bar{\sigma}}$ 
\begin{align}
    \chi_{N}(i \nu_{n}) = \frac{2}{N}\sum_{k,m} v_{k}^{2} G^{\sigma \sigma} (k,i \omega_{m}) G^{\sigma \sigma}(k, i \nu_{n} + i \omega_{m}) 
    \\ \chi_{S}(i \nu_{n}) = \frac{1}{N}\sum_{k,m} v_{k}^{2} G^{\sigma \sigma} (k,i \omega_{m}) G^{\sigma \sigma}(k, i \nu_{n} + i \omega_{m}) 
\end{align}
By writing the spectral representation for the Green\textquotesingle s function $G(k, i \omega_{n})$ and doing analytic continuation, we obtain
\begin{align}
    \chi^{R}(\omega) = \frac{1}{N} \sum_{k} \iint \dd \omega_{1} \dd \omega_{2} \frac{\rho_{G}(k,\omega_{1}) \rho_{G}(k,\omega_{2}) v_{k}^{2}}{\omega + \omega_{1} - \omega_{2} + i \eta} \qty {n_{F}(\omega_{1}) - n_{F}(\omega_{2})}
\end{align}
and $\chi^{R}_{N}(\omega) = 2 \chi^{R}(\omega) $, $\chi^{R}_{S}(\omega) = \chi^{R}(\omega) $.

\section{Self energy low temperature behaviour}
\label{AppendixD}
The low temperature  and low frequency behaviour of the scattering function,
$\Gamma(\omega,T)$ is
\begin{eqnarray*}
\Gamma(\omega,T)&=-\Im \Sigma(\omega,T)=\pi \int_{-\infty}^{\infty} \dd y ~ \rho_{G}(\omega-y) ~ \rho_{D}(y) ~ \qty[n_F(y-\omega)+n_B(y)]\\
&\approx \pi \int_{-\infty}^{\infty} \dd y ~ \rho_{G}(-y) ~ \rho_{D}(y) ~ \qty[n_F(y)+\frac{\omega^2}{2}n_F^{''}(y)+n_B(y)]=I_{1}+I_{2}
\end{eqnarray*}
where
\begin{align*}
I_{1} = 
\pi \int_{-\infty}^{\infty} \dd y ~ \rho_{G}(-y) & ~ \rho_{D}(y) ~ \qty[n_F(y)+n_B(y)] \\
=\pi \int_{0}^{\infty} \dd y ~ \qty(\rho_{G}(-y)+\rho_G(y)) & ~ \rho_{D}(y) ~ \qty(n_F(y)+n_B(y))
\end{align*}

$\rho_D(y)$ is an odd function. We can expand around $y=0$, $\rho_D(y)=Ay$, we obtain

\begin{align}
I_{1} = 2A\pi\rho_{G}(0)\int_{0}^{\infty} \dd y ~ y ~\qty(n_F(y)+n_B(y))=\frac{\pi}{2}A\rho_{G}(0) \pi^{2} T{^2}
\end{align}
And 
\begin{eqnarray}
I_{2} & = -\frac{\pi \omega^2}{2} \int_{-\infty}^{\infty} \dd y ~ \rho_{G}(-y) ~ \rho_{D}(y) ~ \qty(-n_F^{''}(y))=\frac{\pi \omega^2}{2} A \rho_{G}(0) 
\end{eqnarray}

$\Gamma(\omega,T)$ at low temperature and low frequency is 
\begin{eqnarray}
\Gamma(\omega,T)& =\frac{\pi}{2}  \rho_{G}(0)A \qty(\pi^{2} T^{2} + \omega^2)
\end{eqnarray}

\end{document}